\documentclass[journal=jacsat,manuscript=article]{achemso}

\usepackage[version=3]{mhchem} 
\usepackage{amsmath}
\usepackage{amssymb}

\usepackage{subfigure}
\usepackage{xcolor}

\definecolor{green}{RGB}{0,153,51}

\author{Weilong Chen}
\affiliation
{Professorship of Multiscale Modeling of Fluid Materials, Department of Engineering Physics and Computation, TUM School of Engineering and Design, Technical University of Munich, 80333 Munich, Germany}
\author{Franz G\"orlich}
\affiliation
{Professorship of Multiscale Modeling of Fluid Materials, Department of Engineering Physics and Computation, TUM School of Engineering and Design, Technical University of Munich, 80333 Munich, Germany}
\author{Paul Fuchs}
\affiliation
{Professorship of Multiscale Modeling of Fluid Materials, Department of Engineering Physics and Computation, TUM School of Engineering and Design, Technical University of Munich, 80333 Munich, Germany}
\author{Julija Zavadlav}
\email{julija.zavadlav@tum.de}
\affiliation
{Professorship of Multiscale Modeling of Fluid Materials, Department of Engineering Physics and Computation, TUM School of Engineering and Design, Technical University of Munich, 80333 Munich, Germany}
\alsoaffiliation
{Atomistic Modeling Center (AMC), Munich Data Science Institute (MDSI), Technical University of Munich, 85748 Garching, Germany}

\title
  {Enhanced Sampling for Efficient Learning of Coarse-Grained Machine Learning Potentials}

\abbreviations{Machine Learning, Molecular Simulations, Coarse-grained Molecular Dynamics, Enhanced Sampling methods}

\begin{document}

\begin{tocentry}





\end{tocentry}

\begin{abstract}
Coarse-graining (CG) enables molecular dynamics (MD) simulations of larger systems and longer timescales that are otherwise infeasible with atomistic models. Machine learning potentials (MLPs), with their capacity to capture many-body interactions, can provide accurate approximations of the potential of mean force (PMF) in CG models. Current CG MLPs are typically trained in a bottom-up manner via force matching, which in practice relies on configurations sampled from the unbiased equilibrium Boltzmann distribution to ensure thermodynamic consistency. This convention poses two key limitations: first,  sufficiently long atomistic trajectories are needed to reach convergence; and second, even once equilibrated, transition regions remain poorly sampled. To address these issues, we employ enhanced sampling to bias along CG degrees of freedom for data generation, and then recompute the forces with respect to the unbiased potential. This strategy simultaneously shortens the simulation time required to produce equilibrated data and enriches sampling in transition regions, while preserving the correct PMF. We demonstrate its effectiveness on the Müller–Brown potential and capped alanine, achieving notable improvements. Our findings support the use of enhanced sampling for force matching as a promising direction to improve the accuracy and reliability of CG MLPs.
\end{abstract}

\section{Introduction}
Molecular Dynamics (MD) simulations play an important role in science and engineering, providing access to a wide range of structural, dynamical, and thermodynamic properties of molecular systems~\cite{lindorff2011fast, dror2012biomolecular}. In statistical mechanics, such observables can be expressed in terms of expectation values with respect to a statistical distribution (\textit{ensemble}) over microscopic states, defined by macroscopic control parameters such as temperature, volume, pressure, and chemical potential\cite{chandler}. For example, the canonical Boltzmann distribution, $p(\mathbf{r}) = \mathcal{Z}^{-1} \exp\left(-{u(\mathbf{r})}/{k_BT}\right)$ describes the $NVT$ ensemble, in which temperature ($T$), volume ($V$), and particle number ($N$) remain constant. For molecular systems, the high dimensionality of configuration space makes direct evaluation of the partition function  $\mathcal{Z}$  intractable. In practice, sampling based methods such as Markov Chain Monte Carlo (MCMC) or MD\cite{understandingmd} are used to generate configurations from the target distribution. However, the rugged free energy landscapes characteristic of many molecular systems lead to slow decorrelation between samples, making it necessary to run prohibitively long simulations to obtain independent statistics. As a result, extensive sampling of large macromolecular complexes on relevant timescales remains beyond the reach of atomistic resolution.

To address this challenge, a variety of enhanced sampling methods have been developed~\cite{Henin2022}. These approaches accelerate the exploration of the configuration space either by modifying the statistical ensemble to promote rapid transitions between free energy basins or by coupling simulations across multiple thermodynamic ensembles (\textit{replicas})~\cite{chipotFreeEnergyCalculations2007, Marinari_1992}. A notable family of approaches are biasing methods, which perform importance sampling by applying a bias potential that can be reweighted to recover unbiased ensemble statistics~\cite{Ferrenberg_1989,0801.1426}. The bias potential can be static, as in umbrella sampling~\cite{Torrie1977}, or updated dynamically, as in metadynamics~\cite{Laio2002}, and is typically defined in terms of a small set of collective variables that capture the slow degrees of freedom of the system.

Coarse-graining (CG) offers a complementary approach by simplifying atomistic models into reduced representations that capture essential interactions.~\cite{noidPerspectiveCoarsegrainedModels2013, Noid2023, noidRigorousProgressCoarseGraining2024a}. This reduction in dimensionality smooths the free energy surface and decreases computational cost, thus extending the time and length scales accessible to simulation~\cite{marrinkMARTINIForceField2007} while also improving the statistical efficiency of the reweighting procedures. In many applications~\cite{bernardi2015enhanced, yang2019enhanced}, enhanced sampling and coarse-graining can be combined, allowing researchers to combine the benefits of both methods to efficiently explore complex molecular systems.~\cite{klepeis2009long}.

Traditional CG models have historically been parameterized using either "top-down" or "bottom-up" approaches. In top-down approaches, the model parameters are adjusted to reproduce macroscopic observables, such as experimental measurements. A well-known example is the MARTINI model, which was designed to reproduce experimental partition coefficients~\cite{marrinkMARTINIForceField2007}. In bottom-up approaches~\cite{jinBottomupCoarseGrainingPrinciples2022}, the primary goal is to construct a CG model that reproduces the equilibrium configurational distribution (or free energy landscape) of the fine-grained system~\cite{noidPerspectiveCoarsegrainedModels2013}. While this ensures the model captures correct static correlations, it does not guarantee the accurate reproduction of dynamical properties.~\cite{jinBottomupCoarseGrainingPrinciples2022,noidPerspectiveCoarsegrainedModels2013}. By construction, the exact CG potential is the many-body potential of mean force (PMF). However, traditional attempts to approximate the PMF using functional forms or large basis sets similar to classical all-atom potentials have generally proven limited, as added complexity rarely guarantees improved accuracy or transferability~\cite{noidPerspectiveCoarsegrainedModels2013}.

Deep learning has opened new avenues for equilibrium sampling of CG systems~\cite{jing2025ai}. An important direction is the development of CG machine learning potentials (MLPs)~\cite{durumericMachineLearnedCoarsegrained2023, john2017many}, which aim to learn the CG PMF~\cite{zhangDeePCGConstructingCoarsegrained2018, ml-coarse-grain}, analogous to the potential energy function in atomistic systems. These models are typically trained using bottom-up approaches such as variational force matching~\cite{charron2025navigating, Majewski2023} (FM) or relative entropy minimization~\cite{Noid2008}. In FM, the model is trained to minimize the mean squared error between the predicted CG forces and the atomistic forces projected onto the CG space. However, MLPs often depend on prior potentials to ensure reliable predictions outside the training domain, and the corresponding free energy surface is sensitive to errors in transition regions. Relative entropy minimization provides an alternative, but is computationally more expensive due to the requirement of repeated simulations during training~\cite{shell2008relative, thalerDeepCoarsegrainedPotentials2022}. Recent work has also focused on improving the accuracy and transferability of CG MLPs~\cite{thalerLearningNeuralNetwork2021, chenMachineLearningImplicit2021, husicCoarseGrainingMolecular2020, duschatkoThermodynamicallyInformedMultimodal2024, wang2025many, shinkle2024thermodynamic, duschatko2024uncertainty, mondal2025graph}. 

Another active line of research explores deep generative models for CG systems~\cite{noe2019boltzmann, schreiner2023implicit, tamagnoneCoarseGrainedMolecularDynamics2024a, WangBombarelli2019,costa2025morphology,fu2022simulate, hummerich2025split}. Boltzmann Emulators~\cite{ lewis2025scalable,jing2024alphafold, zheng2024predicting}, for example, act as surrogate models by learning a biased distribution that enables one-shot sampling. The connection between generative models and molecular dynamics has led to new sampling approaches. For instance, Flow-Matching~\cite{kohler2023flow} improves data efficiency by training a normalizing flow to approximate the target distribution and then derives forces from the generated samples to train a CG MLP. This shares the goal of relative entropy minimization in reproducing the target distribution, but circumvents the need for iterative CG simulations. Diffusion models provide another approach by directly estimating forces via the score function to enable CG MD simulations~\cite{Arts2023, durumericLearningDataEfficient2024, máté2024neuralthermodynamicintegrationfree, nagelFokkerPlanckScoreLearning2025, plainerConsistentSamplingSimulation2025}. Despite these advances, generative CG models face limitations: the lack of an explicit energy function prevents unbiased reweighting, scaling to larger systems remains difficult~\cite{tanScalableEquilibriumSampling2025, tanAmortizedSamplingTransferable2025, transbgs, moqvist2025thermodynamic, schebek2025scalable, diezTransferableGenerativeModels2025}, and training generally requires unbiased CG MD data. Energy-based models offer an alternative~\cite{nam2025enhancing}, since they do not rely on training samples, but typically require a reliable energy predictor~\cite{havensAdjointSamplingHighly2025a, stuppEnergyBasedCoarseGrainingMolecular2025, dern2025energy,kim2025scalable, midgley2022flow}, which in practice depends on the availability of existing CG MLPs~\cite{thalerDeepCoarsegrainedPotentials2022}.

In this work, we revisit a central limitation of variational force matching for coarse-graining: the mean force can only be approximated statistically through microscopic forces with large fluctuation.~\cite{jinBottomupCoarseGrainingPrinciples2022}. In practice, FM relies on long unbiased trajectories, which are computationally demanding and yield samples concentrated around metastable states, with insufficient coverage of transition regions~\cite{noidPerspectiveCoarsegrainedModels2013, noidRigorousProgressCoarseGraining2024a, zhangDeePCGConstructingCoarsegrained2018, ml-coarse-grain, charron2025navigating}. Consequently, even highly flexible CG potentials trained using FM may perform poorly outside the stable basins and struggle to capture the correct relative probabilities of metastable states~\cite{thalerDeepCoarsegrainedPotentials2022, durumericMachineLearnedCoarsegrained2023}. To overcome these challenges, we introduce enhanced sampling methods for efficient data generation. We show that applying a bias along coarse-grained coordinates and recomputing forces with respect to the unbiased atomistic potential leaves the conditional mean force unchanged. This permits training directly on biased trajectories (without reweighting), substantially accelerating convergence while also improving coverage of transition states. We demonstrate the effectiveness of this approach on the Müller–Brown potential and capped alanine solvated in explicit water. Taken together, our results establish enhanced sampling as a powerful and general framework for constructing accurate and data-efficient CG MLPs, offering fundamental improvements over existing methods.
 
\section{Theory and Methods}
 The coarse-grained modeling begins with the definition of a mapping from the atomistic (AT) description to a reduced set of CG variables. Denote the AT coordinates as $\mathbf{r} \in \mathbb{R}^{3n}$ and the CG coordinates as $\mathbf{R} = \xi(\mathbf{r}) \in \mathbb{R}^{3N}$, with $N < n$. The mapping $\xi$ groups atoms into beads, reducing dimensionality while providing a basis for constructing effective interactions that reproduce microscopic behavior. In this work, we focus on equilibrium thermodynamics in the canonical ($NVT$) ensemble and assume a linear and orthogonal mapping. Extensions to nonlinear mappings, non-equilibrium systems, and kinetic modeling have also been studied~\cite{Nske2019, yang2023slicing, nateghi2025kinetically}.

To preserve the equilibrium distribution, the central requirement for a CG model is thermodynamic consistency: the equilibrium distribution of the CG system must reproduce the equilibrium distribution of the underlying AT system projected onto the CG variables. For canonical ensemble, the AT equilibrium distribution is given by 
\begin{equation}
    p_\text{AT}(\mathbf{r}) = \mathcal{Z}^{-1} \exp\!\left(-\frac{u(\mathbf{r})}{k_BT}\right),
\end{equation}
where $u(\mathbf{r})$ is the AT potential, $k_B$ is the Boltzmann constant, $T$ is the temperature, and $\mathcal{Z} = \int \exp\!\left(-u(\mathbf{r})/k_BT\right) d\mathbf{r}$ is the partition function. The CG equilibrium distribution is obtained by marginalizing over the atomistic degrees of freedom,
\begin{equation}
    p_\text{CG}(\mathbf{R}) = \int \delta(\mathbf{R} - \xi(\mathbf{r}))\, p_\text{AT}(\mathbf{r}) \, d\mathbf{r}.
\end{equation}
The exact many-body potential of mean force (PMF) $U^*(\mathbf{R})$ is defined by the relation:
\begin{equation}
    U^*(\mathbf{R}) = -k_B T \ln p_\text{CG}(\mathbf{R}) + C,
\end{equation}
where $C$ is an arbitrary additive constant. Since exact marginalization is generally intractable for complex systems, a parametric model $U(\mathbf{R}; \theta)$ is typically learned to approximate the exact PMF. The parameters $\theta$ are determined by minimizing a variational objective, such as force matching or relative entropy minimization, such that $U(\mathbf{R}; \theta) \approx U^*(\mathbf{R})$.

\subsection{Force Matching} 
Variational force matching, also known as multiscale coarse-graining~\cite{Noid2008}, is a commonly used approach to learn the CG MLP $U(\mathbf{R};\theta)$. The central idea is that the CG forces predicted by the model should match the instantaneous atomistic forces projected onto the CG coordinates, $\xi(\mathbf{f}(\mathbf{r}))$. The FM loss is defined as the mean squared error between the projected AT forces and the predicted CG forces:
\begin{equation}
    \chi^2(\theta) = \Big\langle \big\| \, \xi(\mathbf{f}(\mathbf{r})) + \nabla U(\xi(\mathbf{r}); \theta) \, \big\|^2 \Big\rangle_{\mathbf{r}},
    \label{eqn:fm_loss}
\end{equation}
where the average is taken over the equilibrium AT distribution.

The FM loss can be further decomposed into two terms~\cite{Noid2008, ml-coarse-grain,zhangDeePCGConstructingCoarsegrained2018},
\begin{equation}
    \chi^2(\theta) = 
    \underbrace{\Big\langle \big\| \mathbf{F}(\mathbf{R}) + \nabla U(\mathbf{R}; \theta) \big\|^2 \Big\rangle_\mathbf{R}}_{\text{PMF error}}
    \;+\; 
    \underbrace{\text{Noise}(\xi)}_{\text{irreducible}},
\end{equation}
where $\mathbf{F}(\mathbf{R}) = \langle \xi(\mathbf{f}(\mathbf{r})) \rangle_{\mathbf{r}|\mathbf{R}}$ is the \textit{mean force} conditioned on the CG state. The first term measures the deviation between the mean force $\mathbf{F}(\mathbf{R})$ and the CG forces predicted by the CG potential. The second term, $\text{Noise}(\xi)$, represents the irreducible variance of the projected atomistic forces arising from the many-to-one nature of the mapping $\xi$. This noise term depends only on the choice of mapping and cannot be reduced by optimizing the CG model. Hence, the machine learning task in FM is to find a potential $U(\mathbf{R};\theta)$ that best approximates the mean force $\mathbf{F}(\mathbf{R})$. For this reason, $U$ is often referred to as the \textit{potential of mean force} (PMF). Minimizing $\chi^2(\theta)$ ensures that the learned potential approximates the PMF as closely as possible given the chosen CG mapping and available data. 
In practice, given a finite dataset of $M$ atomistic configurations $\mathcal{D}=\{\mathbf{r}_1,\ldots,\mathbf{r}_M\}$, the empirical FM loss can be estimated as
\begin{equation}
    \hat{\chi}^2(\theta) 
    = \frac{1}{3M} \sum_{i=1}^M
    \Big\| \, \xi(\mathbf{f}(\mathbf{r}_i)) + \nabla U(\xi(\mathbf{r}_i);\theta) \, \Big\|^2,
\end{equation}
where $\xi(\mathcal{D}) = [\xi(\mathbf{r}_1), \ldots, \xi(\mathbf{r}_M)]^\top \in \mathbb{R}^{M\times 3N}$ and $\xi(\mathbf{f}(\mathcal{D})) = [\xi(\mathbf{f}(\mathbf{r}_1)), \ldots, \xi(\mathbf{f}(\mathbf{r}_M))]^\top \in \mathbb{R}^{M\times 3N}$.

\subsection{Finite Data Size Effects}\label{subsec:finite_data}
Learning CG MLPs under the force matching (FM) framework is fundamentally limited by \textit{finite data size effects}. Two main factors contribute to this challenge.

The first arises from the nature of CG force matching. Unlike atomistic MLPs, where potential energy labels are directly available, the CG PMF $U(\mathbf{R})$ must be inferred indirectly from instantaneous forces by minimizing the variational bound in Eq.~\ref{eqn:fm_loss}. The true mean force $\mathbf{F}(\mathbf{R})$ is defined as the average of all atomistic configurations corresponding to the same coarse-grained state $\mathbf{R}$, while a single projected force represents only one noisy sample of this average. Accurate approximation of this mean requires dense sampling in the neighborhood of each $\mathbf{R}$, so that statistical noise does not dominate the learning signal. As a result, FM-trained CG models generally require much larger datasets than atomistic models trained on explicit energy surfaces. Although conditional averages could, in principle, be obtained using constrained MD or Blue Moon sampling~\cite{Carter1989, ciccotti2005blue}, performing such targeted sampling for a sufficiently dense set of coarse-grained configurations is computationally prohibitive.

The second factor stems from how CG FM datasets are generated in practice. To ensure thermodynamic consistency, configurations are typically sampled from the unbiased equilibrium Boltzmann distribution of the atomistic system. Producing sufficiently long atomistic trajectories is necessary to achieve convergence of the mean forces, which can be particularly challenging for complex biomolecular systems due to rare events and slow transitions~\cite{lindorff2011fast}. Even once equilibrium is reached, the samples are unevenly distributed: Configurations are concentrated near metastable states, whereas transition regions remain poorly represented. This uneven sampling reduces the accuracy of the mean-force approximation in less-populated regions of configuration space.

\subsection{Unbiased Mean Forces from Biased Sampling}
As we describe, a major limitation of CG force matching is the practical difficulty of generating sufficiently representative equilibrium data. Transition regions, rarely visited in standard MD, are particularly underrepresented, resulting in noisy mean force estimates and less reliable CG MLPs. Enhanced sampling methods, such as umbrella sampling, metadynamics, or other biasing strategies, are natural choices to improve coverage of these regions. A natural question arises: does training on biased data distort the CG mean force and thereby compromise thermodynamic consistency?

We formalize this question as follows. Let $W(\mathbf{R})$ denote a bias potential applied along the CG coordinates. The biased AT distribution is given by
\begin{equation}
    p_W(\mathbf{r}) \;=\; \mathcal{Z}_W^{-1} \exp\!\left(-\beta \big( u(\mathbf{r}) + W(\xi(\mathbf{r})) \big) \right),
\end{equation}
The key observation is that although the bias changes the marginal distribution of the CG coordinates, it does not alter the conditional distribution of atomistic configurations at fixed $\mathbf{R}$.
The conditional distribution under the bias is
\begin{equation}
    p_W(\mathbf{r} \mid \mathbf{R}) 
    \;=\; \frac{\delta(\xi(\mathbf{r}) - \mathbf{R}) \, e^{-\beta \big( u(\mathbf{r}) + W(\mathbf{R}) \big)}}{\int \delta(\xi(\mathbf{r}) - \mathbf{R}) \, e^{-\beta \big( u(\mathbf{r}) + W(\mathbf{R}) \big)} \, d\mathbf{r}}.
\end{equation}
Since $W(\mathbf{R})$ is constant when conditioning on $\mathbf{R}$, it cancels between numerator and denominator:
\begin{equation}
    p_W(\mathbf{r} \mid \mathbf{R}) \;=\; 
    \frac{\delta(\xi(\mathbf{r}) - \mathbf{R}) \, e^{-\beta u(\mathbf{r})}}{\int \delta(\xi(\mathbf{r}) - \mathbf{R}) \, e^{-\beta u(\mathbf{r})} \, d\mathbf{r}} 
    \;=\; p(\mathbf{r} \mid \mathbf{R}),
\end{equation}
Thus, any bias that depends exclusively on the CG variables leaves the atomistic conditional ensemble at fixed $\mathbf{R}$ unchanged.
Consequently, the coarse-grained mean force, defined as the conditional expectation of the projected forces, remains invariant:
\begin{equation}
    \mathbf{F}(\mathbf{R}) 
    \;=\; \big\langle \, \xi(\mathbf{f}(\mathbf{r})) \, \big\rangle_{\mathbf{r}\,|\, \mathbf{R}}
    \;=\; \big\langle \, \xi(\mathbf{f}(\mathbf{r})) \, \big\rangle_{\mathbf{r_W}\,|\, \mathbf{R}}.
    \label{eq:biased_invariance}
\end{equation}
Intuitively, the bias alters how frequently a given CG configuration $\mathbf{R}$ is visited, but once $\mathbf{R}$ is fixed, the conditional distribution of atomistic microstates consistent with $\mathbf{R}$ is unaffected by $W$. In practice, however, the invariance derived in Eq.~\ref{eq:biased_invariance} relies on several underlying assumptions, outlined below:
        \begin{enumerate}
            \item \textbf{Functional Dependence of the Bias:} The bias potential must depend exclusively on the CG variables, i.e., $W = W(\xi(\mathbf{r}))$. Dependence on orthogonal degrees of freedom would perturb the conditional ensemble $p(\mathbf{r}|\mathbf{R})$, rendering the sampled mean forces thermodynamically inconsistent. In such cases, explicit reweighting is required to recover unbiased mean forces.
        
            \item \textbf{Conditional Ergodicity:} The system needs to remain ergodic in the orthogonal degrees of freedom. While the bias accelerates the exploration of $\mathbf{R}$, the fast degrees of freedom (e.g., vibrations, solvent) are assumed to equilibrate rapidly to the conditional Boltzmann distribution. Hysteresis or trapping in these hidden variables would yield non-equilibrium force estimates.
        
            \item \textbf{Recovery of Unbiased Forces:} The training targets must be the forces derived from the unbiased potential $u(\mathbf{r})$. The biasing force contribution, $-\nabla_{\mathbf{r}} W(\xi(\mathbf{r}))$, must be subtracted from biased forces.
        \end{enumerate}

\begin{figure}
    \centering
    \includegraphics[width=0.8\linewidth]{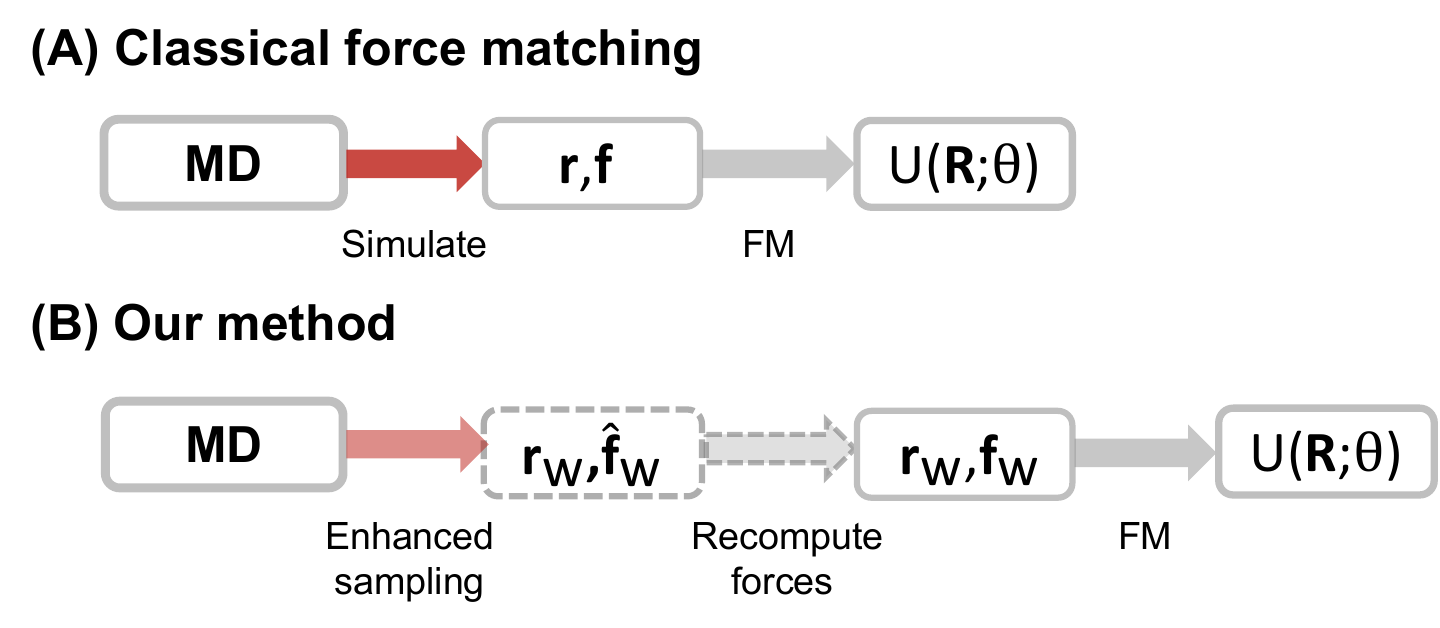}
    \caption{\textbf{Overview of the enhanced sampling force matching method.}
    (A) Classical force matching: positions $\mathbf{r}$ and forces $\mathbf{f}$ from an unbiased atomistic MD simulation are used to learn the potential of mean force (PMF) $U$.  
    (B) Enhanced sampling force matching (this work): Configurations are obtained via
    enhanced sampling, reducing the required simulation time (light red region). The
    forces acting during the biased simulation are
    $\hat{\mathbf{f}}_W(\mathbf{r}_W) = -\nabla_{\mathbf{r}}\!\left(u(\mathbf{r}_W) +
    W(\xi(\mathbf{r}_W))\right)$,
    while the unbiased forces used for training are recomputed using the unbiased potential,
    $\mathbf{f}_W(\mathbf{r}_W) = -\nabla_{\mathbf{r}} u(\mathbf{r}_W)$,
    which incurs minimal additional computational cost. The PMF is learned from the
    biased configurations $\mathbf{r}_W$ and their corresponding recomputed forces
    $\mathbf{f}_W$.
    }
    \label{fig:workflow}
\end{figure}

Subject to the assumptions above, the optimization over the biased distribution targets the identical mean force as the unbiased case. The biased objective function is defined as:
\begin{equation}
    \chi^2_W(\theta) 
    \;=\; \Big\langle \big\| \, \xi(\mathbf{f}(\mathbf{r})) + \nabla U(\xi(\mathbf{r}); \theta) \, \big\|^2 \Big\rangle_{\mathbf{r_W}}.
    \label{eq:biased_loss}
\end{equation}
While the scalar value of the loss $\chi^2_W$ differs from the unbiased objective $\chi^2$ due to the modified sampling density, both functionals share the same global minimum with respect to $\theta$ (see SI for a detailed proof). This invariance enables the training of CG potentials on datasets generated with biased sampling, without requiring reweighting of the loss function. The practical advantages are twofold: (i) biased simulations accelerate exploration of rarely visited states, reducing the total simulation time needed for data generation, and (ii) the resulting datasets provide more uniform coverage of both energy basins and transition regions, leading to more accurate and robust CG MLPs. An overview of the enhanced sampling force matching workflow is shown in Figure~\ref{fig:workflow}.

\subsection{Enhanced Sampling Methods}

The invariance of mean forces under CG coordinate-dependent biasing (Eq.~\ref{eq:biased_invariance}) shows that biased simulations can be directly used for force matching, provided that forces are recomputed with respect to the unbiased atomistic potential. This observation allows us to incorporate enhanced sampling methods that accelerate exploration of rarely visited or high-barrier regions. Here, we focus on two popular choices: umbrella sampling and well-tempered metadynamics.

\paragraph{Umbrella sampling.}  
Umbrella sampling~\cite{Torrie1977} improves sampling efficiency by applying a static bias potential $W(\mathbf{R})$ that confines the system near a chosen region of the (CG) coordinate space. In this work, we employ a single harmonic constraint centered on $\mathbf{R}_0$,  
\begin{equation}
    W(\mathbf{R}) = \tfrac{1}{2} \kappa \|\mathbf{R} - \mathbf{R}_0\|^2,
\end{equation}
where $\kappa$ is the force constant. The biased AT distribution is
\begin{equation}
    p_{W}(\mathbf{r}) \;\propto\; \exp\!\Big(-\beta \big(u(\mathbf{r}) + W(\xi(\mathbf{r}))\big)\Big).
\end{equation}
This ensures better sampling of the configurations around $\mathbf{R}_0$, allowing a better representation of transition regions that are otherwise rarely observed in unbiased trajectories.

\paragraph{Well-tempered metadynamics.}  
Metadynamics~\cite{Laio2002} enhances sampling by progressively filling free energy basins with a history-dependent bias, thereby discouraging revisiting previously explored regions. At time intervals $\tau$, Gaussians of width $\sigma$ and initial height $h$ are deposited along the chosen CG (or CV) coordinates,
\begin{equation}
    W_t(\mathbf{R}) = \sum_{\tau < t} h \exp\!\left(-\frac{\|\mathbf{R}-\mathbf{R}(\tau)\|^2}{2\sigma^2}\right).
\end{equation}
In plain metadynamics, the bias keeps growing indefinitely, eventually flattening the free energy surface. Well-Tempered metadynamics~\cite{barducci2008well} improves this by tempering the Gaussian heights with a bias factor $\gamma>1$,  
\begin{equation}
    W_t(\mathbf{R}) = \sum_{\tau < t} h \exp\!\left(-\frac{\|\mathbf{R}-\mathbf{R}(\tau)\|^2}{2\sigma^2}\right)
    \exp\!\left(-\frac{W_\tau(\mathbf{R})}{k_B T (\gamma-1)}\right).
\end{equation}
In the long-time limit, this yields sampling from a modified distribution,
\begin{equation}
    p_{\text{WT}}(\mathbf{R}) \;\propto\; \exp\!\left(-\tfrac{\beta}{\gamma} A(\mathbf{R})\right),
\end{equation}
where $A(\mathbf{R})$ is the free energy surface of the CG (or CV) coordinates R. This corresponds to sampling at an effective temperature $T^* = \gamma T$, since the bias factor $\gamma = T^*/T$ rescales the thermal fluctuations along $\mathbf{R}$. Physically, this implies that the coarse-grained variables explore the landscape as if coupled to a high-temperature heat bath at $T^*$, effectively reducing barrier heights by a factor of $1/\gamma$. Note that the remaining atomistic degrees of freedom (e.g., solvent, fast vibrations) remain at the physical temperature $T$. This allows the system to overcome high-energy barriers associated with the collective variables without inducing the unphysical structural denaturation that would occur in a global high-temperature simulation. The term “well-tempered” reflects the fact that the bias is added more slowly over time, striking a balance between the exploration of new regions and the preservation of meaningful free energy differences.

A practical challenge in applying enhanced sampling to novel systems is selecting bias parameters without prior knowledge of the free-energy landscape. Importantly, our method prioritizes achieving sufficient exploration of the conformational space rather than the exact convergence of the bias potential required for quantitative free-energy reconstruction. This distinction makes the method more robust to sub-optimal parameter choices. For novel systems, we recommend an iterative refinement strategy. Initial parameters can be selected based on standard heuristics~\cite{barducci2008well, bonomi2009plumed, laio2008metadynamics}: for solvated biomolecules, the bias factor is typically chosen in the range $\gamma \in [5, 20]$ to adequately reduce expected barriers, while the Gaussian width $\sigma$ is estimated from natural thermal fluctuations observed in short unbiased precursor runs (e.g., $\sigma \approx 1/3$)~\cite{tribello2025plumed}. Sampling quality is then assessed by monitoring the histogram of the collective variables. If unsampled regions or “gaps” persist along transition pathways, the bias strength can be increased to help bridge these regions.

\subsection{Graph Neural Network Potentials without Priors}
To parameterize the CG potential $U(\mathbf{R};\theta)$ for molecular systems, we adopt the MACE architecture~\cite{batatia2022mace}, an equivariant message-passing graph neural network originally developed for atomistic potentials. Each CG bead is represented as a node in a graph, and edges indicate neighbor pairs within a cutoff radius and carry distance/relative-vector embeddings. Equivariant message passing layers update node features while enforcing that $U(\mathbf{R};\theta)$ is invariant under rigid translations and rotations and that internal vector features transform equivariantly. The force on bead $i$ is obtained directly from the learned potential by automatic differentiation,
\begin{equation}
    \mathbf{F}_i(\mathbf{R};\theta) \;=\; 
    - \nabla_{\mathbf{R}_i} U(\mathbf{R};\theta),
\end{equation}
ensuring that forces are conservative by construction.

A common strategy in the literature is to augment CG MLPs with a physics-based prior, such as baseline pairwise interactions and harmonic bonded terms, so that the network only learns a corrective energy term, which is helpful for data efficiency and improves simulation stability. In contrast, here we train \textsc{MACE} directly on the force-matching loss (Eq.~\ref{eqn:fm_loss}) without including any prior~\cite{mirarchi2024amaro}. This choice avoids introducing modeling bias and allows the network to learn the PMF purely from data.

\section{Results}
We evaluate the performance of our methods by applying them to study two representative systems: a Low-dimensional M\"uller–Brown Potential and MD simulation data of capped alanine in water. More detailed information on both systems can be found in the Supporting Information. For both systems, we apply enhanced sampling by introducing suitable bias potentials to promote exploration of rarely visited states. The biased forces are recomputed at each sampled configuration with respect to the unbiased potential and serve as training data for the CG MLPs. We evaluate the methods in terms of data efficiency and model accuracy. Specific hyperparameter choices for all our experiments can be found in the Supporting Information.

\subsection{Low-dimensional M\"uller Brown Potential}

\begin{figure}[h!]
    \centering
    \includegraphics[width=0.5\linewidth]{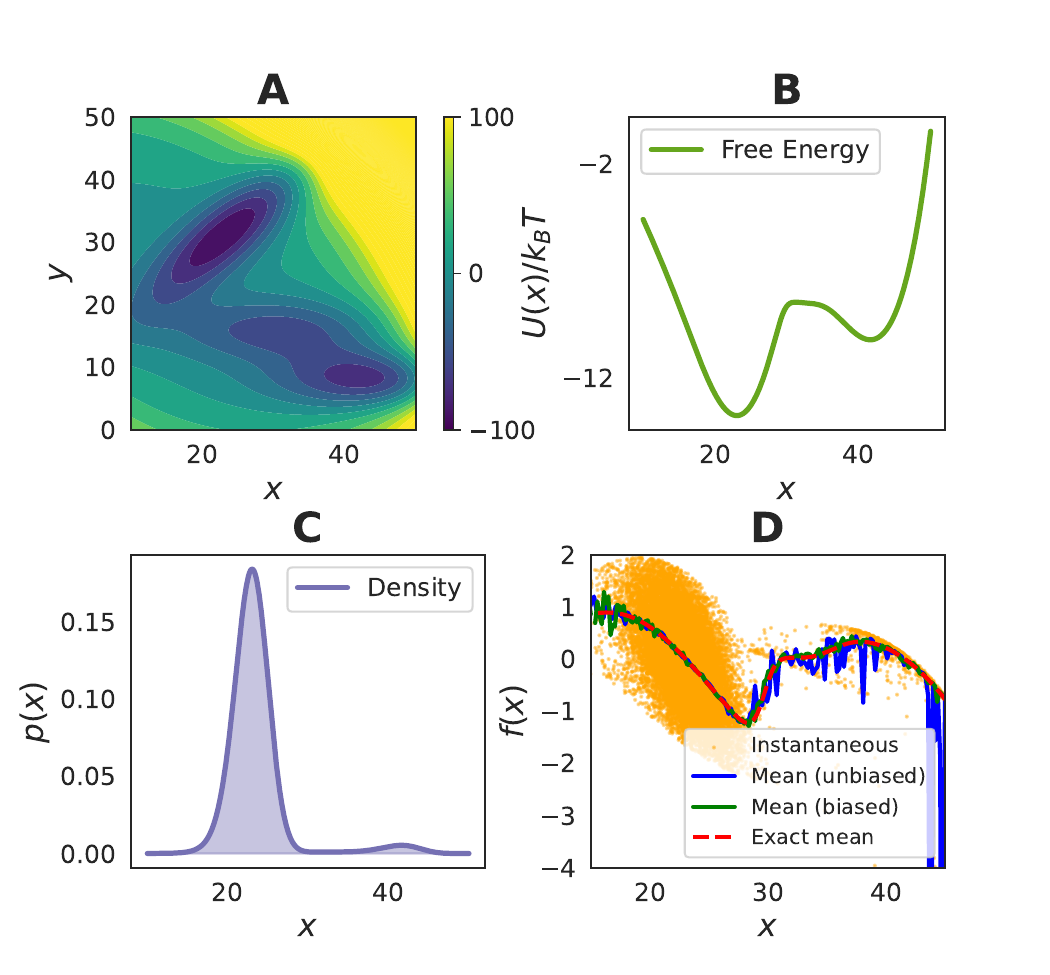}
    \caption{%
    \textbf{Finite data size effects in the low-dimensional M\"uller–Brown system.}  
    (A) Two-dimensional M\"uller–Brown potential energy surface (functional form given in the Supporting Information).  
    (B) Exact free-energy profile along the $x$-axis.  
    (C) Marginal probability density along the $x$-axis.  
    (D) Instantaneous force samples from the unbiased dataset projected onto the $x$-axis, shown together with the exact mean force and bin-averaged estimates from biased and unbiased datasets of equal size. The unit of the force is $k_BT$.
    }
    \label{fig:finite_data}
\end{figure}

We consider the two-dimensional M\"uller–Brown (MB) potential, a canonical test system for transition path sampling~\cite{rajaActionMinimizationMeetsGenerative2025}, which features a global minimum and two local minima separated by saddle points (Figure~\ref{fig:finite_data}A). 
Coarse-graining is defined here as projection onto the $x$-axis. The corresponding CG PMF is given exactly by 
\begin{equation}
    \frac{U(x)}{k_B T} = - \ln \int_{-\infty}^{\infty} \exp(-\beta u(x,y)) \, dy.
\end{equation} with the result shown in Figure~\ref{fig:finite_data}B. The associated probability density along $x$ is $p_\text{CG}(x) = \mathcal{Z}_x^{-1}{\exp(-\beta U(x))},$ as shown in Figure~\ref{fig:finite_data}C.
The exact mean force along $x$ is obtained from the derivative $-{dU(x)}/{dx}$ and is plotted in Figure~\ref{fig:finite_data}D.

To generate training data, we performed two types of simulations. First, an unbiased trajectory is run until equilibrium, sampling the 
Boltzmann distribution 
\(p(\mathbf{r}) \propto \exp(-\beta u(\mathbf{r}))\). 
Second, a biased trajectory is generated using umbrella sampling, with 
a Gaussian restraint \(w(\mathbf{r})\) applied close to the barrier region 
to enhance transitions between metastable basins. 
This simulation samples the biased distribution 
\(p_W(\mathbf{r}) \propto \exp(-\beta (u(\mathbf{r}) + w(\mathbf{r})))\).  

For each configuration \(\mathbf{r}\) generated in either simulation, 
we record the positions \(\mathbf{r}\), the unbiased forces 
\(\mathbf{f}(\mathbf{r}) = - \nabla u(\mathbf{r})\), and the biased 
forces \(\hat{\mathbf{f}}(\mathbf{r}) = - \nabla (u(\mathbf{r}) + w(\mathbf{r}))\). 
In the biased case, we also compute importance weights 
\(\omega(\mathbf{r}) = \exp(\beta w(\mathbf{r}))\), which allow 
reweighting to recover unbiased equilibrium averages. Specifically, any observable \(\phi(\mathbf{r})\) can be estimated by self-normalized importance sampling \(\mathbb{E}_{p}[\phi(\mathbf{r})] \approx \sum_{i=1}^{K} \bar{\omega}(\mathbf{r}_i) \, \phi(\mathbf{r}_i)\), with \(\bar{\omega}(\mathbf{r}_i) = \omega(\mathbf{r}_i)/\sum_{j=1}^{K} \omega(\mathbf{r}_j)\), where the sum runs over the \(K\) configurations sampled from the biased trajectory.

\paragraph{Finite data size effects}
Sampling from the unbiased equilibrium distribution results in a highly uneven coverage: Most configurations accumulate in the left minimum, while other basins are rarely visited (Figure~\ref{fig:finite_data}C). This imbalance is further illustrated in Figure~\ref{fig:finite_data}D, which shows 20,000 instantaneous force samples projected onto the $x$ axis, with bin averages used to approximate the mean force. In regions with dense sampling, such as the left basin, the estimated mean force agrees closely with the exact result. In contrast, poorly sampled regions, particularly the right minimum, yield noisy and inaccurate estimates.

This behavior highlights a general limitation of equilibrium simulations with high-energy barriers: finite datasets provide imbalanced and incomplete coverage, and
force matching suffers as a result. Biased sampling provides a natural solution: as shown in Figure~\ref{fig:finite_data}D, bin-averaged mean forces from biased datasets of equal size recover the correct mean force profile with substantially reduced variance. This empirically demonstrates that enhanced sampling alleviates finite data size effects, a challenge that becomes even more pronounced in higher dimensional systems.

\begin{figure}
    \centering
    \includegraphics[width=0.5\linewidth]{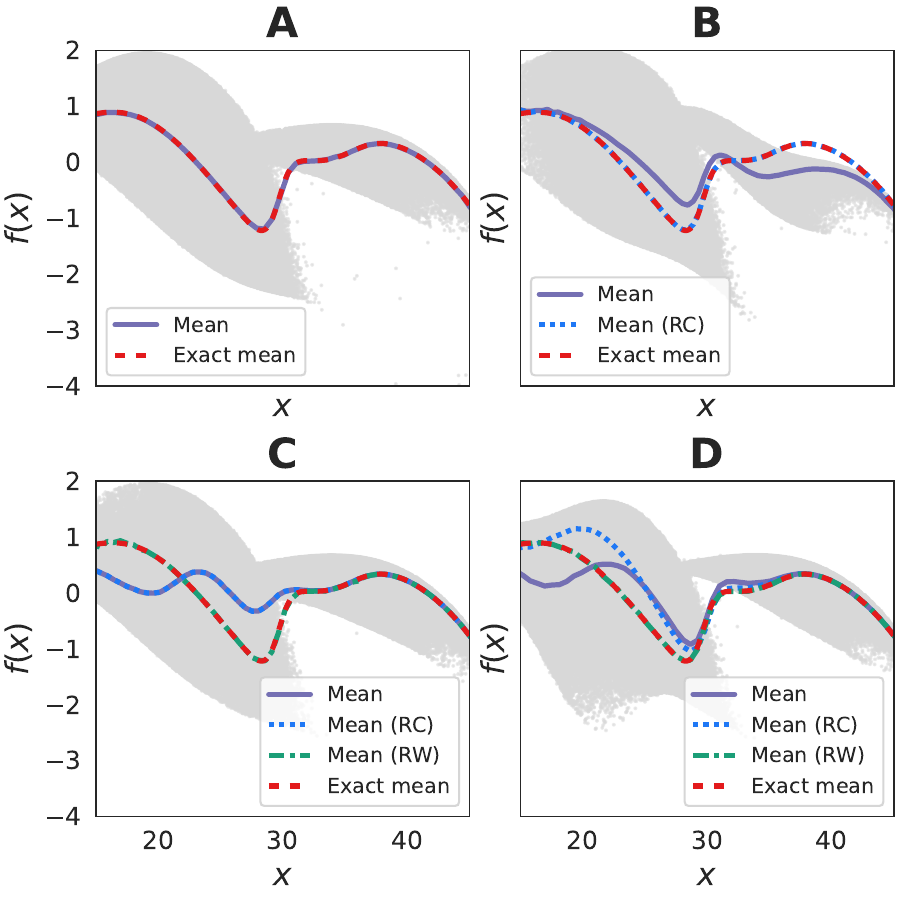}
    \caption{%
    \textbf{Unbiased mean force recovery in the low-dimensional M\"uller–Brown system.}    
    In all panels, gray dots show instantaneous forces from the corresponding simulations.  
    Overlaid curves denote the exact mean force and bin-averaged estimates:  
    (A) Unbiased simulation.  
    (B) Biased along $x$: bin-averaged estimates include both direct and recomputed (RC) mean forces.  
    (C) Biased along $y$: bin-averaged estimates include direct, RC, and reweighted (RW) mean forces from importance sampling.  
    (D) Biased along $(x,y)$: bin-averaged estimates include direct, RC, and RW mean forces.  
    }
    \label{fig:unbiased_mean_force}
\end{figure}

\paragraph{Unbiased mean forces}
We next verify that the recomputed mean force remains unbiased if and only if the bias is applied along the coarse-grained degree of freedom. To this end, we generated three datasets with biasing potentials applied along $x$, $y$, and $(x,y)$, each containing sufficient samples to accurately estimate the mean force (Figure~\ref{fig:unbiased_mean_force}). When the bias is applied only along $x$, the mean force profile along $x$ is correctly recovered after recomputing the forces with respect to the unbiased potential, without the need for reweighting (Figure~\ref{fig:unbiased_mean_force}B). In contrast, when the bias acts along $y$ or jointly along $(x,y)$, reweighting is required to recover the correct mean force (Figure~\ref{fig:unbiased_mean_force}C–D).

Next, we trained machine learning potentials on both unbiased and biased datasets, restricting the bias to the $x$ coordinate, to assess its effect on model accuracy. Potentials are parameterized using a neural network with \emph{radial basis function} (RBF) features as input, followed by several fully connected layers. The RBF layer maps coordinates into a high-dimensional feature space,  
\begin{equation}
    \phi_j(x) = \exp\!\left(-\frac{(x - c_j)^2}{2\sigma^2}\right), \qquad j=1,\dots,K,
\end{equation}  
where $\{c_j\}$ denote the centers and $\sigma$ controls the width of the features. These localized features improve the ability of the network to capture nonlinear variations in the mean force landscape compared to using raw coordinates (Supporting Information Figure~S1). 

\begin{figure}[h!]
    \centering
    \includegraphics[width=1.0\linewidth]{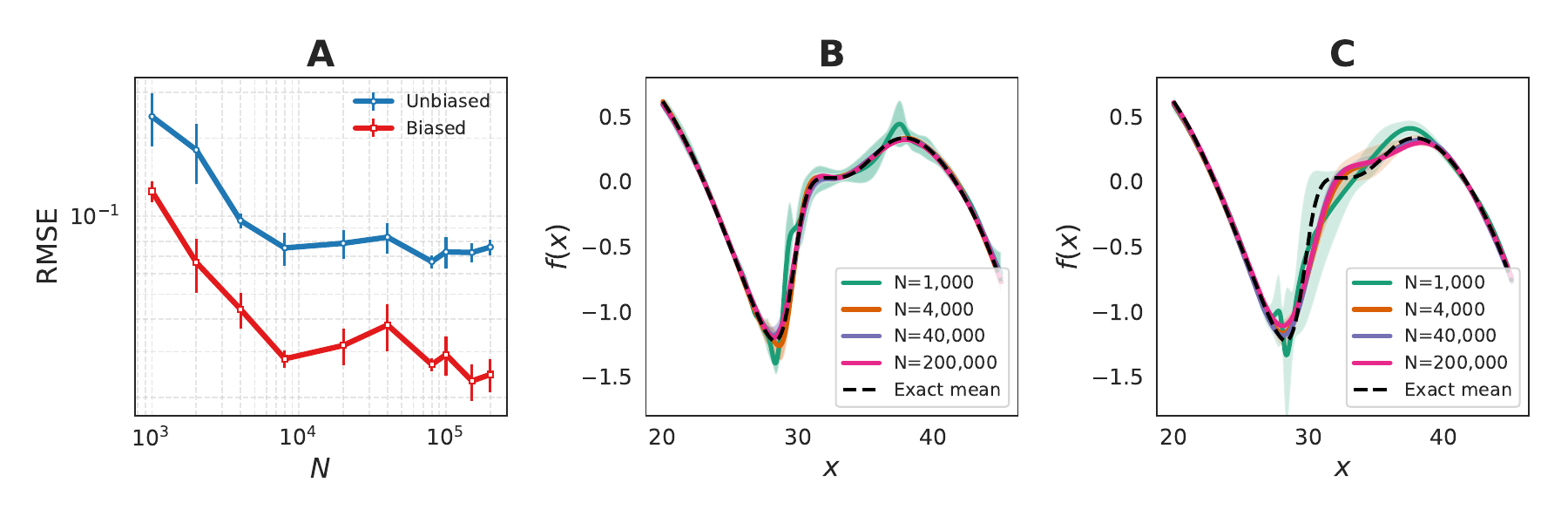}
    \caption{%
    \textbf{Results for the low-dimensional Müller--Brown potential.}
    (A) Root-mean-square error (RMSE) of predicted forces as a function of the number of training samples ($N$). 
    RMSE values are computed relative to the exact mean force over 500 equally spaced points in the interval $x \in [20,45]$. 
    Results are shown for models trained on biased datasets generated with umbrella sampling and on unbiased datasets. 
    Error bars represent the standard deviation across five independently trained models with different random seeds. 
    (B) Exact mean force compared with model-predicted forces trained on biased datasets obtained via umbrella sampling. 
    $N$ indicates the number of training samples; uncertainties reflect variations across five independently trained models. 
    (C) Same as (B), but using unbiased datasets for training.
    }

    \label{fig:mse_vs_data_size}
\end{figure}

Figure~\ref{fig:mse_vs_data_size}A reports the mean-squared error (MSE) between the predicted and exact mean force as a function of the amount of training data. Models trained on biased datasets reach lower error and variance with only a few thousand samples, whereas models trained on unbiased datasets require orders of magnitude more data, yet still exhibit larger variance and higher error. Direct comparison of the learned force curves (Figure~\ref{fig:mse_vs_data_size}B–C) further illustrates this difference: While both models reproduce the mean force in densely sampled regions, biased training achieves much lower uncertainty and accurately recovers both the overall shape and the fine features of the mean force with limited data. Unbiased training, on the contrary, captures only the broad trend and fails to reproduce local structure even with orders of magnitude more samples. We additionally compare the exact free energy profile with MLP simulations in the Supporting Information (Figure~S2), where simulation profiles are obtained via the negative natural logarithm of the sampled position histograms.

\subsection{Coarse-Graining of Capped Alanine in Water.}
\begin{figure}[h!]
    \centering
    \includegraphics[width=1\linewidth]{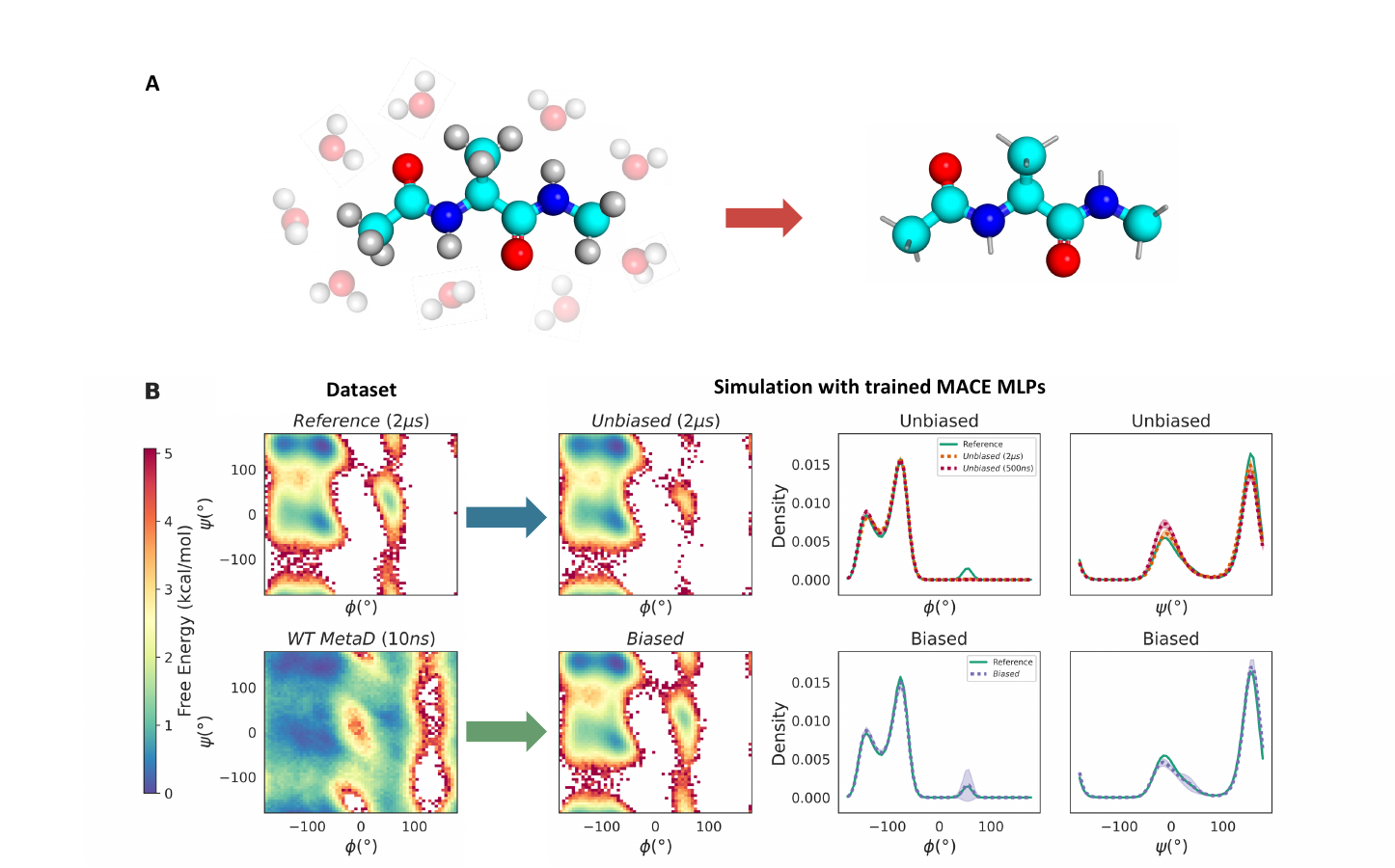} 
    \caption{%
    \textbf{Coarse-graining mapping of capped alanine and the resulting free energy profiles.}
    (A) Mapping from the all-atom solvated model (left) to the coarse-grained (CG) model retaining the ten heavy backbone atoms (right). 
    (B) Free-energy surfaces and one-dimensional dihedral distributions for datasets and CG model simulations. The left column (“Dataset”) shows the reference 2~µs unbiased MD free-energy surface and the well-tempered metadynamics (WT MetaD, 10~ns) dataset used for model training. The right columns (“Simulation with trained MACE MLPs”) show the corresponding free-energy surfaces and one-dimensional $\phi/\psi$ dihedral distributions obtained from CG simulations using models trained on the respective datasets. Mean values and standard deviations (shaded regions) are computed from 100 independent CG trajectories of 100~ns each.
    }
    \label{fig:joint}
\end{figure}

As for the molecular benchmark, we demonstrate our approach on the coarse-graining of solvated capped alanine (alanine dipeptide), a prototypical system for conformational transitions. The coarse-grained mapping retains all ten heavy atoms while discarding hydrogens and water molecules, as illustrated in Figure~\ref{fig:joint}A.

We generate both unbiased and biased datasets for training. The unbiased dataset is obtained from a 500~ns MD trajectory at 300~K, from which $5 \times 10^5$ configurations are sampled uniformly in time. Biased datasets are generated using well-tempered metadynamics (WT MetaD) with backbone dihedrals $\phi$ (C–N–C$\alpha$–C) and $\psi$ (N–C$\alpha$–C–N) as collective variables, employing PLUMED~\cite{bonomi2009plumed} with GROMACS~\cite{van2005gromacs}. Simulations are performed with bias factors $\gamma = 1.5, 3, 6, 9$, each of length 10~ns, and $5 \times 10^5$ samples are collected per dataset. The Ramachandran plots corresponding to these datasets are shown in the Supporting Information (Figure~S6). As $\gamma$ increases, the simulations explore progressively larger regions of conformational space, particularly transition regions between metastable basins. To explicitly validate the quality of the sampling, we employed two practical convergence criteria: (i) the time-evolution of the biased collective variables to ensure rapid mixing between states, and (ii) the stability of the reconstructed free energy profiles over time across independent replicas. As detailed in the Supporting Information (Figures S3--S4), the biased simulations exhibit frequent transitions and thermodynamic convergence (stabilization of $\Delta F$) within few nanoseconds. In contrast, the unbiased reference simulation fails to achieve comparable convergence even after 500~ns due to rare transition events. A sufficiently long 2~µs unbiased MD trajectory is generated as a reference. For all biased simulations, instantaneous forces are recomputed with respect to the unbiased potential using the \texttt{rerun} feature of GROMACS.

To illustrate the effects of finite data size and mean force invariance in the molecular system, we consider a generalized coordinate \(q\), such as a dihedral angle of the backbone \(\theta\). The conjugate force is  
\begin{equation}
    Q_\theta = \sum_i \mathbf{f}_i \cdot \frac{\partial \mathbf{r}_i}{\partial \theta},
\end{equation}
where $\mathbf{f}_i = -\partial u/\partial \mathbf{r}_i$ is the Cartesian force on atom $i$ and $\partial \mathbf{r}_i / \partial \theta$ is its displacement under a unit change in $\theta$. $Q_\theta$ represents the generalized torque that drives the rotation around the dihedral. As shown in the SI Figure~S5, mean generalized torque \(\langle Q_\theta \rangle\) calculated from unbiased trajectories fluctuates strongly in sparsely sampled transition regions, illustrating the limitations of equilibrium data in capturing the full conformational landscape. In contrast, recomputing forces from biased trajectories yields mean torques with much lower variance and correctly recovers the reference profile, confirming invariance under CG coordinate-dependent bias (Eq.~\ref{eq:biased_invariance}).  

We then train the MACE model on these datasets using the \texttt{chemtrain} framework~\cite{fuchs2025chemtraina, fuchs2025chemtrainb}, with the same hyperparameter settings (listed in the SI). CG simulations are performed under Langevin dynamics at 300~K using JAX~M.D.~\cite{schoenholz2020jax}. For evaluation, we run $100$ independent CG simulations of $100$~ns each, initialized from random configurations, for both the unbiased dataset and biased datasets with different bias factors \(\gamma\). Ramachandran plots and Dihedral distributions (Figure~\ref{fig:joint}B) show that models trained on unbiased data fail to recover the metastable basin $\alpha_\mathrm{L}$ at $\phi \approx 0^\circ$–$100^\circ$ on the right-hand side of the Ramachandran map, whereas biased training with sufficiently large $\gamma$ recovers both modes accurately. Quantifying metastable populations across five independent models (Supporting Information Figure~S7-8) show that unbiased datasets and low-$\gamma$ WT MetaD assign nearly zero probability to the metastable state $\alpha_\mathrm{L}$, while higher-$\gamma$ datasets accurately capture it. 

\begin{figure}[h!]
    \centering
    \includegraphics[width=1.0\linewidth]{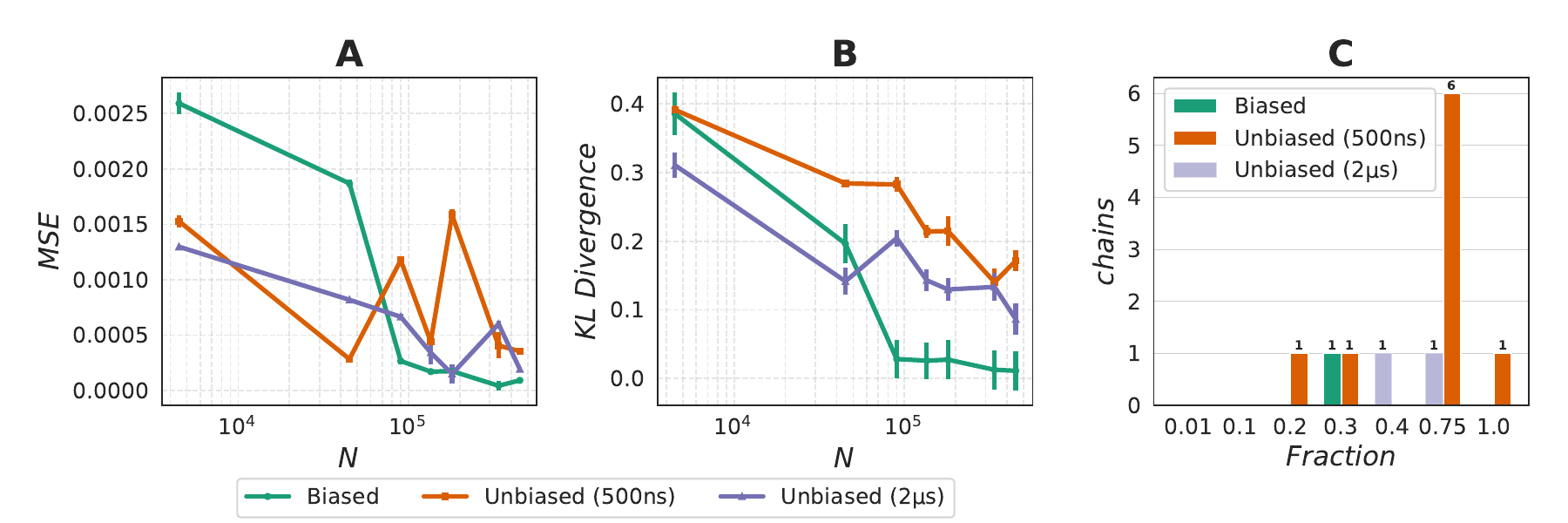} 
    \caption{
    \textbf{Model accuracy and stability for capped alanine.} 
    (A) Mean squared error (MSE) between discrete free energies on the $\phi/\psi$ plane for varying training data sizes. 
    Mean and standard deviation are estimated from 100 trajectories of 100~ns each. 
    (B) Kullback--Leibler (KL) divergence between discrete free energies on the $\phi/\psi$ plane for varying training data sizes. 
    Mean and standard deviation are again computed from 100 trajectories of 100~ns each. 
    (C) Number of unstable trajectories as a function of training data size, expressed as a fraction of the total samples. 
    Numbers indicate how many out of 100 trajectories are unstable; if not specified, all trajectories are stable.
    }
    \label{fig:mse_kl}
\end{figure}

Next, we investigate the effect of the size of the training dataset. For each, we run 100 independent 100~ns CG simulations and compare the resulting $\phi$–$\psi$ distributions to reference MD. Specifically, we compute the KL divergence and mean-squared error (MSE) of the torsional free energy on discrete histograms (Figure~\ref{fig:mse_kl}). 
For MSE, unbiased datasets initially yield smaller errors as a result of denser sampling of the left-hand mode in the Ramachandran map. However, as the size of the dataset increases, biased simulations achieve a lower overall MSE by accurately reproducing both modes. For KL divergence, biased datasets consistently outperform the 500~ns unbiased dataset, and surpass the 2~µs unbiased dataset when more samples are used, as they capture the global free-energy landscape more faithfully. Together, these results highlight the better accuracy of training on biased datasets.

Finally, we assess stability by monitoring numerical instabilities across training fractions (Figure~\ref{fig:mse_kl}C). Chains are considered unstable and removed when the predicted potential energy reaches unphysically high values. In previous analyses, these unstable chains were already excluded; here we explicitly report their occurrence. Most simulations remain stable across 100~ns, with failures occurring only in a few chains. For the 500~ns unbiased dataset, up to six chains diverge at fraction 0.75, with isolated failures at fractions 0.2, 0.3, and 1.0. The 2~µs unbiased dataset shows single-chain failures at fractions 0.4 and 0.75. In contrast, biased datasets exhibit only one failure at fraction 0.3. These results indicate that biased datasets improve both accuracy and stability by providing broader coverage of configuration space. For completeness, we report free energy surfaces without chain removal as well as per-chain results (SI, Figure~S9-10). Additionally, to assess computational efficiency, we present a wall-clock time comparison between our MACE model and the reference ATMD simulations performed in GROMACS (SI, Table~A2).

\section{Conclusion}
Our work introduces enhanced sampling as a principled strategy for generating training data and improving the efficiency of training CG MLPs within the force matching framework. We show that mean forces are invariant under biases applied along CG degrees of freedom once the forces are recomputed, enabling biased trajectories to be used directly for training without reweighting. Using umbrella sampling and well-tempered metadynamics as representative enhanced sampling methods, we demonstrate on both the Müller–Brown potential and capped alanine that biased datasets provide substantially improved force coverage and data efficiency, yielding accurate and stable CG models without the need for physics-based priors.

Our results demonstrate that enhanced sampling provides a practical solution to the finite data size effects inherent in force matching. By accelerating transitions across energy barriers, enhanced sampling significantly reduces data generation time. It also enriches the training dataset with configurations that are rarely visited in unbiased simulations. As a result, neural networks can reconstruct the potential of mean force with higher accuracy, particularly in transition regions. Notably, this improvement is achieved without introducing additional physical priors: enhanced sampling itself supplies the necessary regularization. In this way, it functions as a data-side regularizer, allowing complex CG interactions to be learned directly from data, while reducing the dependence on hand-crafted corrections.

One limitation of our approach is its dependence on prior knowledge of collective variables (CVs) or reaction coordinates suitable for biasing. In the Müller–Brown and capped alanine benchmarks, the relevant slow modes are well understood, allowing the bias to be applied directly to the coarse-grained degrees of freedom. However, for more complex biomolecular systems, it is challenging to identify such CVs~\cite{wang2019past,noe2013variational,perez2013identification}. When chosen CVs do not adequately capture the slow dynamics, enhanced sampling may not sufficiently enrich force coverage, limiting the improvement of the resulting models. Machine learning techniques for automated reaction coordinate discovery~\cite{zhang2024flow,ribeiro2018reweighted,chen2018molecular,herringer2023permutationally,mehdi2024enhanced} provide a potential solution, and their integration could facilitate a broader application to high-dimensional CG mappings and larger biomolecules.

Looking ahead, the framework allows for several natural extensions. Biasing could be applied not only along predefined collective variables, but also along arbitrary coarse-grained degrees of freedom, including learned slow coordinates. Additional enhanced sampling methods, such as adaptive biasing force~\cite{darve2001calculating,gao2008self,zhang2018reinforced}, replica exchange or tempering~\cite{Marinari_1992}, variationally enhanced sampling~\cite{valsson2014variational, piaggi2019multithermal}, could readily be integrated to further improve efficiency and transferability. Furthermore, one could relax the requirement of biasing only the mapped coordinates by applying bias to orthogonal degrees of freedom; this would necessitate recovering unbiased mean forces via explicit reweighting strategies~\cite{wang2025marginal, bause2021reweighting} (e.g., via Markov state models~\cite{prinz2011markov, bause2019microscopic}). Such extensions share conceptual roots with established frameworks for transferring potentials between thermodynamic states~\cite{krishna2009multiscale, jin2020temperature, shinkle2024thermodynamic, wu2024structural}. However, the benefit of biasing orthogonal degrees of freedom is likely limited in this context, since coarse-grained mappings are typically designed to explicitly capture the slowest, most relevant coordinates.

Additionally, an active learning cycle alternating between model training, uncertainty quantification of mean forces, and targeted bias placement would enable systematic sampling of regions with high uncertainty, producing datasets that are both efficient and informative~\cite{duschatko2024uncertainty,vitartas2025active, kulichenko2023uncertainty}. The balance between biased and unbiased simulations can also be optimized, for example, by employing pretraining-finetuning paradigms that take advantage of complementary data sources~\cite{rocken2025enhancing}. From a practical perspective, our approach could be used to construct or refine large-scale datasets for training transferable CG MLPs~\cite{sillitoe2015cath,mirarchi2024mdcath}, improving transferability across varying thermodynamic conditions and chemical compositions. It could also provide information for generative models that typically lack force supervision~\cite{wang2024protein} or support energy-based neural samplers~\cite{liu2025adjoint,havensAdjointSamplingHighly2025a, he2025no}. Overall, we believe that our method represents a fundamental advance over current methodologies and opens new opportunities to tackle outstanding challenges for efficient learning of coarse-gained molecular models.

\suppinfo
Details of dataset generation, training procedures, hyperparameters, computational cost, and additional qualitative results on molecular systems are available in the Supporting Information.

\begin{acknowledgement}
Funded by the European Union. Views and opinions expressed are however those of the author(s) only and do not necessarily reflect those of the European Union or the European Research Council Executive Agency. Neither the European Union nor the granting authority can be held responsible for them. This work was funded by the ERC (StG SupraModel) - 101077842 and the Deutsche Forschungsgemeinschaft (DFG, German Research Foundation) - 534045056 and 561190767. 

\end{acknowledgement}


\section*{Data and Code Availability}
The code and data supporting this study will be made publicly available on GitHub upon acceptance of this manuscript.

\bibliography{ref}

\providecommand{\latin}[1]{#1}
\makeatletter
\providecommand{\doi}
  {\begingroup\let\do\@makeother\dospecials
  \catcode`\{=1 \catcode`\}=2 \doi@aux}
\providecommand{\doi@aux}[1]{\endgroup\texttt{#1}}
\makeatother
\providecommand*\mcitethebibliography{\thebibliography}
\csname @ifundefined\endcsname{endmcitethebibliography}
  {\let\endmcitethebibliography\endthebibliography}{}
\begin{mcitethebibliography}{110}
\providecommand*\natexlab[1]{#1}
\providecommand*\mciteSetBstSublistMode[1]{}
\providecommand*\mciteSetBstMaxWidthForm[2]{}
\providecommand*\mciteBstWouldAddEndPuncttrue
  {\def\EndOfBibitem{\unskip.}}
\providecommand*\mciteBstWouldAddEndPunctfalse
  {\let\EndOfBibitem\relax}
\providecommand*\mciteSetBstMidEndSepPunct[3]{}
\providecommand*\mciteSetBstSublistLabelBeginEnd[3]{}
\providecommand*\EndOfBibitem{}
\mciteSetBstSublistMode{f}
\mciteSetBstMaxWidthForm{subitem}{(\alph{mcitesubitemcount})}
\mciteSetBstSublistLabelBeginEnd
  {\mcitemaxwidthsubitemform\space}
  {\relax}
  {\relax}

\bibitem[Lindorff-Larsen \latin{et~al.}(2011)Lindorff-Larsen, Piana, Dror, and
  Shaw]{lindorff2011fast}
Lindorff-Larsen,~K.; Piana,~S.; Dror,~R.~O.; Shaw,~D.~E. How fast-folding
  proteins fold. \emph{Science} \textbf{2011}, \emph{334}, 517--520\relax
\mciteBstWouldAddEndPuncttrue
\mciteSetBstMidEndSepPunct{\mcitedefaultmidpunct}
{\mcitedefaultendpunct}{\mcitedefaultseppunct}\relax
\EndOfBibitem
\bibitem[Dror \latin{et~al.}(2012)Dror, Dirks, Grossman, Xu, and
  Shaw]{dror2012biomolecular}
Dror,~R.~O.; Dirks,~R.~M.; Grossman,~J.; Xu,~H.; Shaw,~D.~E. Biomolecular
  simulation: a computational microscope for molecular biology. \emph{Annual
  review of biophysics} \textbf{2012}, \emph{41}, 429--452\relax
\mciteBstWouldAddEndPuncttrue
\mciteSetBstMidEndSepPunct{\mcitedefaultmidpunct}
{\mcitedefaultendpunct}{\mcitedefaultseppunct}\relax
\EndOfBibitem
\bibitem[Chandler(1987)]{chandler}
Chandler,~D. \emph{Introduction to Modern Statistical Mechanics}; Oxford
  University Press, 1987\relax
\mciteBstWouldAddEndPuncttrue
\mciteSetBstMidEndSepPunct{\mcitedefaultmidpunct}
{\mcitedefaultendpunct}{\mcitedefaultseppunct}\relax
\EndOfBibitem
\bibitem[Frenkel and Smit(2002)Frenkel, and Smit]{understandingmd}
Frenkel,~D.; Smit,~B. \emph{Understanding Molecular Simulation}; Elsevier,
  2002\relax
\mciteBstWouldAddEndPuncttrue
\mciteSetBstMidEndSepPunct{\mcitedefaultmidpunct}
{\mcitedefaultendpunct}{\mcitedefaultseppunct}\relax
\EndOfBibitem
\bibitem[H{\'e}nin \latin{et~al.}(2022)H{\'e}nin, Leli{\`e}vre, Shirts,
  Valsson, and Delemotte]{Henin2022}
H{\'e}nin,~J.; Leli{\`e}vre,~T.; Shirts,~M.~R.; Valsson,~O.; Delemotte,~L.
  Enhanced Sampling Methods for Molecular Dynamics Simulations [Article v1.0].
  \emph{Living Journal of Computational Molecular Science} \textbf{2022},
  \emph{4}\relax
\mciteBstWouldAddEndPuncttrue
\mciteSetBstMidEndSepPunct{\mcitedefaultmidpunct}
{\mcitedefaultendpunct}{\mcitedefaultseppunct}\relax
\EndOfBibitem
\bibitem[Chipot and Pohorille(2007)Chipot, and
  Pohorille]{chipotFreeEnergyCalculations2007}
Chipot,~C.; Pohorille,~A. \emph{Free energy calculations}; Springer, 2007;
  Vol.~86\relax
\mciteBstWouldAddEndPuncttrue
\mciteSetBstMidEndSepPunct{\mcitedefaultmidpunct}
{\mcitedefaultendpunct}{\mcitedefaultseppunct}\relax
\EndOfBibitem
\bibitem[Marinari and Parisi(1992)Marinari, and Parisi]{Marinari_1992}
Marinari,~E.; Parisi,~G. Simulated Tempering: A New Monte Carlo Scheme.
  \emph{Europhysics Letters (EPL)} \textbf{1992}, \emph{19}, 451–458\relax
\mciteBstWouldAddEndPuncttrue
\mciteSetBstMidEndSepPunct{\mcitedefaultmidpunct}
{\mcitedefaultendpunct}{\mcitedefaultseppunct}\relax
\EndOfBibitem
\bibitem[Ferrenberg and Swendsen(1989)Ferrenberg, and
  Swendsen]{Ferrenberg_1989}
Ferrenberg,~A.~M.; Swendsen,~R.~H. Optimized Monte Carlo data analysis.
  \emph{Physical Review Letters} \textbf{1989}, \emph{63}, 1195--1198\relax
\mciteBstWouldAddEndPuncttrue
\mciteSetBstMidEndSepPunct{\mcitedefaultmidpunct}
{\mcitedefaultendpunct}{\mcitedefaultseppunct}\relax
\EndOfBibitem
\bibitem[Shirts and Chodera(2008)Shirts, and Chodera]{0801.1426}
Shirts,~M.~R.; Chodera,~J.~D. Statistically optimal analysis of samples from
  multiple equilibrium states. \emph{The Journal of Chemical Physics}
  \textbf{2008}, \emph{129}\relax
\mciteBstWouldAddEndPuncttrue
\mciteSetBstMidEndSepPunct{\mcitedefaultmidpunct}
{\mcitedefaultendpunct}{\mcitedefaultseppunct}\relax
\EndOfBibitem
\bibitem[Torrie and Valleau(1977)Torrie, and Valleau]{Torrie1977}
Torrie,~G.; Valleau,~J. Nonphysical sampling distributions in Monte Carlo
  free-energy estimation: Umbrella sampling. \emph{Journal of Computational
  Physics} \textbf{1977}, \emph{23}, 187--199\relax
\mciteBstWouldAddEndPuncttrue
\mciteSetBstMidEndSepPunct{\mcitedefaultmidpunct}
{\mcitedefaultendpunct}{\mcitedefaultseppunct}\relax
\EndOfBibitem
\bibitem[Laio and Parrinello(2002)Laio, and Parrinello]{Laio2002}
Laio,~A.; Parrinello,~M. {Escaping free-energy minima}. \emph{Proceedings of
  the National Academy of Sciences} \textbf{2002}, \emph{99},
  12562--12566\relax
\mciteBstWouldAddEndPuncttrue
\mciteSetBstMidEndSepPunct{\mcitedefaultmidpunct}
{\mcitedefaultendpunct}{\mcitedefaultseppunct}\relax
\EndOfBibitem
\bibitem[Noid(2013)]{noidPerspectiveCoarsegrainedModels2013}
Noid,~W.~G. Perspective: {{Coarse-grained}} Models for Biomolecular Systems.
  \emph{The Journal of Chemical Physics} \textbf{2013}, \emph{139},
  090901\relax
\mciteBstWouldAddEndPuncttrue
\mciteSetBstMidEndSepPunct{\mcitedefaultmidpunct}
{\mcitedefaultendpunct}{\mcitedefaultseppunct}\relax
\EndOfBibitem
\bibitem[Noid(2023)]{Noid2023}
Noid,~W.~G. Perspective: Advances, Challenges, and Insight for Predictive
  Coarse-Grained Models. \emph{The Journal of Physical Chemistry B}
  \textbf{2023}, \emph{127}, 4174–4207\relax
\mciteBstWouldAddEndPuncttrue
\mciteSetBstMidEndSepPunct{\mcitedefaultmidpunct}
{\mcitedefaultendpunct}{\mcitedefaultseppunct}\relax
\EndOfBibitem
\bibitem[Noid \latin{et~al.}(2024)Noid, Szukalo, Kidder, and
  Lesniewski]{noidRigorousProgressCoarseGraining2024a}
Noid,~W.; Szukalo,~R.~J.; Kidder,~K.~M.; Lesniewski,~M.~C. Rigorous
  {{Progress}} in {{Coarse-Graining}}. \emph{Annual Review of Physical
  Chemistry} \textbf{2024}, \emph{75}, 21--45\relax
\mciteBstWouldAddEndPuncttrue
\mciteSetBstMidEndSepPunct{\mcitedefaultmidpunct}
{\mcitedefaultendpunct}{\mcitedefaultseppunct}\relax
\EndOfBibitem
\bibitem[Marrink \latin{et~al.}(2007)Marrink, Risselada, Yefimov, Tieleman, and
  {de Vries}]{marrinkMARTINIForceField2007}
Marrink,~S.~J.; Risselada,~H.~J.; Yefimov,~S.; Tieleman,~D.~P.; {de
  Vries},~A.~H. The {{MARTINI Force Field}}:\, {{Coarse Grained Model}} for
  {{Biomolecular Simulations}}. \emph{The Journal of Physical Chemistry B}
  \textbf{2007}, \emph{111}, 7812--7824\relax
\mciteBstWouldAddEndPuncttrue
\mciteSetBstMidEndSepPunct{\mcitedefaultmidpunct}
{\mcitedefaultendpunct}{\mcitedefaultseppunct}\relax
\EndOfBibitem
\bibitem[Bernardi \latin{et~al.}(2015)Bernardi, Melo, and
  Schulten]{bernardi2015enhanced}
Bernardi,~R.~C.; Melo,~M.~C.; Schulten,~K. Enhanced sampling techniques in
  molecular dynamics simulations of biological systems. \emph{Biochimica et
  Biophysica Acta (BBA)-General Subjects} \textbf{2015}, \emph{1850},
  872--877\relax
\mciteBstWouldAddEndPuncttrue
\mciteSetBstMidEndSepPunct{\mcitedefaultmidpunct}
{\mcitedefaultendpunct}{\mcitedefaultseppunct}\relax
\EndOfBibitem
\bibitem[Yang \latin{et~al.}(2019)Yang, Shao, Zhang, Yang, and
  Gao]{yang2019enhanced}
Yang,~Y.~I.; Shao,~Q.; Zhang,~J.; Yang,~L.; Gao,~Y.~Q. Enhanced sampling in
  molecular dynamics. \emph{The Journal of chemical physics} \textbf{2019},
  \emph{151}\relax
\mciteBstWouldAddEndPuncttrue
\mciteSetBstMidEndSepPunct{\mcitedefaultmidpunct}
{\mcitedefaultendpunct}{\mcitedefaultseppunct}\relax
\EndOfBibitem
\bibitem[Klepeis \latin{et~al.}(2009)Klepeis, Lindorff-Larsen, Dror, and
  Shaw]{klepeis2009long}
Klepeis,~J.~L.; Lindorff-Larsen,~K.; Dror,~R.~O.; Shaw,~D.~E. Long-timescale
  molecular dynamics simulations of protein structure and function.
  \emph{Current opinion in structural biology} \textbf{2009}, \emph{19},
  120--127\relax
\mciteBstWouldAddEndPuncttrue
\mciteSetBstMidEndSepPunct{\mcitedefaultmidpunct}
{\mcitedefaultendpunct}{\mcitedefaultseppunct}\relax
\EndOfBibitem
\bibitem[Jin \latin{et~al.}(2022)Jin, Pak, Durumeric, Loose, and
  Voth]{jinBottomupCoarseGrainingPrinciples2022}
Jin,~J.; Pak,~A.~J.; Durumeric,~A. E.~P.; Loose,~T.~D.; Voth,~G.~A. Bottom-up
  {{Coarse-Graining}}: {{Principles}} and {{Perspectives}}. \emph{Journal of
  Chemical Theory and Computation} \textbf{2022}, \emph{18}, 5759--5791\relax
\mciteBstWouldAddEndPuncttrue
\mciteSetBstMidEndSepPunct{\mcitedefaultmidpunct}
{\mcitedefaultendpunct}{\mcitedefaultseppunct}\relax
\EndOfBibitem
\bibitem[Jing \latin{et~al.}(2025)Jing, Berger, and Jaakkola]{jing2025ai}
Jing,~B.; Berger,~B.; Jaakkola,~T. AI-based Methods for Simulating, Sampling,
  and Predicting Protein Ensembles. \emph{arXiv preprint arXiv:2509.17224}
  \textbf{2025}, \relax
\mciteBstWouldAddEndPunctfalse
\mciteSetBstMidEndSepPunct{\mcitedefaultmidpunct}
{}{\mcitedefaultseppunct}\relax
\EndOfBibitem
\bibitem[Durumeric \latin{et~al.}(2023)Durumeric, Charron, Templeton, Musil,
  Bonneau, {Pasos-Trejo}, Chen, Kelkar, No{\'e}, and
  Clementi]{durumericMachineLearnedCoarsegrained2023}
Durumeric,~A. E.~P.; Charron,~N.~E.; Templeton,~C.; Musil,~F.; Bonneau,~K.;
  {Pasos-Trejo},~A.~S.; Chen,~Y.; Kelkar,~A.; No{\'e},~F.; Clementi,~C. Machine
  Learned Coarse-Grained Protein Force-Fields: {{Are}} We There Yet?
  \emph{Current Opinion in Structural Biology} \textbf{2023}, \emph{79},
  102533\relax
\mciteBstWouldAddEndPuncttrue
\mciteSetBstMidEndSepPunct{\mcitedefaultmidpunct}
{\mcitedefaultendpunct}{\mcitedefaultseppunct}\relax
\EndOfBibitem
\bibitem[John and Cs{\'a}nyi(2017)John, and Cs{\'a}nyi]{john2017many}
John,~S.; Cs{\'a}nyi,~G. Many-body coarse-grained interactions using Gaussian
  approximation potentials. \emph{The Journal of Physical Chemistry B}
  \textbf{2017}, \emph{121}, 10934--10949\relax
\mciteBstWouldAddEndPuncttrue
\mciteSetBstMidEndSepPunct{\mcitedefaultmidpunct}
{\mcitedefaultendpunct}{\mcitedefaultseppunct}\relax
\EndOfBibitem
\bibitem[Zhang \latin{et~al.}(2018)Zhang, Han, Wang, Car, and
  E]{zhangDeePCGConstructingCoarsegrained2018}
Zhang,~L.; Han,~J.; Wang,~H.; Car,~R.; E,~W. {{DeePCG}}: {{Constructing}}
  Coarse-Grained Models via Deep Neural Networks. \emph{The Journal of Chemical
  Physics} \textbf{2018}, \emph{149}, 034101\relax
\mciteBstWouldAddEndPuncttrue
\mciteSetBstMidEndSepPunct{\mcitedefaultmidpunct}
{\mcitedefaultendpunct}{\mcitedefaultseppunct}\relax
\EndOfBibitem
\bibitem[Wang \latin{et~al.}(2019)Wang, Olsson, Wehmeyer, Pérez, Charron,
  de~Fabritiis, Noé, and Clementi]{ml-coarse-grain}
Wang,~J.; Olsson,~S.; Wehmeyer,~C.; Pérez,~A.; Charron,~N.~E.;
  de~Fabritiis,~G.; Noé,~F.; Clementi,~C. Machine Learning of Coarse-Grained
  Molecular Dynamics Force Fields. \emph{ACS Central Science} \textbf{2019},
  \emph{5}, 755–767\relax
\mciteBstWouldAddEndPuncttrue
\mciteSetBstMidEndSepPunct{\mcitedefaultmidpunct}
{\mcitedefaultendpunct}{\mcitedefaultseppunct}\relax
\EndOfBibitem
\bibitem[Charron \latin{et~al.}(2025)Charron, Bonneau, Pasos-Trejo, Guljas,
  Chen, Musil, Venturin, Gusew, Zaporozhets, Kr{\"a}mer, \latin{et~al.}
  others]{charron2025navigating}
Charron,~N.~E.; Bonneau,~K.; Pasos-Trejo,~A.~S.; Guljas,~A.; Chen,~Y.;
  Musil,~F.; Venturin,~J.; Gusew,~D.; Zaporozhets,~I.; Kr{\"a}mer,~A.; others
  Navigating protein landscapes with a machine-learned transferable
  coarse-grained model. \emph{Nature Chemistry} \textbf{2025}, 1--9\relax
\mciteBstWouldAddEndPuncttrue
\mciteSetBstMidEndSepPunct{\mcitedefaultmidpunct}
{\mcitedefaultendpunct}{\mcitedefaultseppunct}\relax
\EndOfBibitem
\bibitem[Majewski \latin{et~al.}(2023)Majewski, Pérez, Th\"{o}lke, Doerr,
  Charron, Giorgino, Husic, Clementi, Noé, and De~Fabritiis]{Majewski2023}
Majewski,~M.; Pérez,~A.; Th\"{o}lke,~P.; Doerr,~S.; Charron,~N.~E.;
  Giorgino,~T.; Husic,~B.~E.; Clementi,~C.; Noé,~F.; De~Fabritiis,~G. Machine
  learning coarse-grained potentials of protein thermodynamics. \emph{Nature
  Communications} \textbf{2023}, \emph{14}\relax
\mciteBstWouldAddEndPuncttrue
\mciteSetBstMidEndSepPunct{\mcitedefaultmidpunct}
{\mcitedefaultendpunct}{\mcitedefaultseppunct}\relax
\EndOfBibitem
\bibitem[Noid \latin{et~al.}(2008)Noid, Chu, Ayton, Krishna, Izvekov, Voth,
  Das, and Andersen]{Noid2008}
Noid,~W.~G.; Chu,~J.-W.; Ayton,~G.~S.; Krishna,~V.; Izvekov,~S.; Voth,~G.~A.;
  Das,~A.; Andersen,~H.~C. The multiscale coarse-graining method. I. A rigorous
  bridge between atomistic and coarse-grained models. \emph{The Journal of
  Chemical Physics} \textbf{2008}, \emph{128}\relax
\mciteBstWouldAddEndPuncttrue
\mciteSetBstMidEndSepPunct{\mcitedefaultmidpunct}
{\mcitedefaultendpunct}{\mcitedefaultseppunct}\relax
\EndOfBibitem
\bibitem[Shell(2008)]{shell2008relative}
Shell,~M.~S. The relative entropy is fundamental to multiscale and inverse
  thermodynamic problems. \emph{The Journal of chemical physics} \textbf{2008},
  \emph{129}\relax
\mciteBstWouldAddEndPuncttrue
\mciteSetBstMidEndSepPunct{\mcitedefaultmidpunct}
{\mcitedefaultendpunct}{\mcitedefaultseppunct}\relax
\EndOfBibitem
\bibitem[Thaler \latin{et~al.}(2022)Thaler, Stupp, and
  Zavadlav]{thalerDeepCoarsegrainedPotentials2022}
Thaler,~S.; Stupp,~M.; Zavadlav,~J. Deep Coarse-Grained Potentials via Relative
  Entropy Minimization. \emph{The Journal of Chemical Physics} \textbf{2022},
  \emph{157}, 244103\relax
\mciteBstWouldAddEndPuncttrue
\mciteSetBstMidEndSepPunct{\mcitedefaultmidpunct}
{\mcitedefaultendpunct}{\mcitedefaultseppunct}\relax
\EndOfBibitem
\bibitem[Thaler and Zavadlav(2021)Thaler, and
  Zavadlav]{thalerLearningNeuralNetwork2021}
Thaler,~S.; Zavadlav,~J. Learning neural network potentials from experimental
  data via Differentiable Trajectory Reweighting. \emph{Nature communications}
  \textbf{2021}, \emph{12}, 6884\relax
\mciteBstWouldAddEndPuncttrue
\mciteSetBstMidEndSepPunct{\mcitedefaultmidpunct}
{\mcitedefaultendpunct}{\mcitedefaultseppunct}\relax
\EndOfBibitem
\bibitem[Chen \latin{et~al.}(2021)Chen, Kr{\"a}mer, Charron, Husic, Clementi,
  and No{\'e}]{chenMachineLearningImplicit2021}
Chen,~Y.; Kr{\"a}mer,~A.; Charron,~N.~E.; Husic,~B.~E.; Clementi,~C.;
  No{\'e},~F. Machine Learning Implicit Solvation for Molecular Dynamics.
  \emph{The Journal of Chemical Physics} \textbf{2021}, \emph{155},
  084101\relax
\mciteBstWouldAddEndPuncttrue
\mciteSetBstMidEndSepPunct{\mcitedefaultmidpunct}
{\mcitedefaultendpunct}{\mcitedefaultseppunct}\relax
\EndOfBibitem
\bibitem[Husic \latin{et~al.}(2020)Husic, Charron, Lemm, Wang, P{\'e}rez,
  Majewski, Kr{\"a}mer, Chen, Olsson, De~Fabritiis, No{\'e}, and
  Clementi]{husicCoarseGrainingMolecular2020}
Husic,~B.~E.; Charron,~N.~E.; Lemm,~D.; Wang,~J.; P{\'e}rez,~A.; Majewski,~M.;
  Kr{\"a}mer,~A.; Chen,~Y.; Olsson,~S.; De~Fabritiis,~G.; No{\'e},~F.;
  Clementi,~C. Coarse Graining Molecular Dynamics with Graph Neural Networks.
  \emph{The Journal of Chemical Physics} \textbf{2020}, \emph{153},
  194101\relax
\mciteBstWouldAddEndPuncttrue
\mciteSetBstMidEndSepPunct{\mcitedefaultmidpunct}
{\mcitedefaultendpunct}{\mcitedefaultseppunct}\relax
\EndOfBibitem
\bibitem[Duschatko \latin{et~al.}(2024)Duschatko, Fu, Owen, Xie, Musaelian,
  Jaakkola, and Kozinsky]{duschatkoThermodynamicallyInformedMultimodal2024}
Duschatko,~B.~R.; Fu,~X.; Owen,~C.; Xie,~Y.; Musaelian,~A.; Jaakkola,~T.;
  Kozinsky,~B. Thermodynamically informed multimodal learning of
  high-dimensional free energy models in molecular coarse graining. \emph{arXiv
  preprint arXiv:2405.19386} \textbf{2024}, \relax
\mciteBstWouldAddEndPunctfalse
\mciteSetBstMidEndSepPunct{\mcitedefaultmidpunct}
{}{\mcitedefaultseppunct}\relax
\EndOfBibitem
\bibitem[Wang \latin{et~al.}(2025)Wang, Csanyi, and Ortner]{wang2025many}
Wang,~Y.; Csanyi,~G.; Ortner,~C. Many-Body Coarse-Grained Molecular Dynamics
  with the Atomic Cluster Expansion. \emph{arXiv preprint arXiv:2502.04661}
  \textbf{2025}, \relax
\mciteBstWouldAddEndPunctfalse
\mciteSetBstMidEndSepPunct{\mcitedefaultmidpunct}
{}{\mcitedefaultseppunct}\relax
\EndOfBibitem
\bibitem[Shinkle \latin{et~al.}(2024)Shinkle, Pachalieva, Bahl, Matin, Gifford,
  Craven, and Lubbers]{shinkle2024thermodynamic}
Shinkle,~E.; Pachalieva,~A.; Bahl,~R.; Matin,~S.; Gifford,~B.; Craven,~G.~T.;
  Lubbers,~N. Thermodynamic transferability in coarse-grained force fields
  using graph neural networks. \emph{Journal of Chemical Theory and
  Computation} \textbf{2024}, \emph{20}, 10524--10539\relax
\mciteBstWouldAddEndPuncttrue
\mciteSetBstMidEndSepPunct{\mcitedefaultmidpunct}
{\mcitedefaultendpunct}{\mcitedefaultseppunct}\relax
\EndOfBibitem
\bibitem[Duschatko \latin{et~al.}(2024)Duschatko, Vandermause, Molinari, and
  Kozinsky]{duschatko2024uncertainty}
Duschatko,~B.~R.; Vandermause,~J.; Molinari,~N.; Kozinsky,~B. Uncertainty
  driven active learning of coarse grained free energy models. \emph{npj
  Computational Materials} \textbf{2024}, \emph{10}, 9\relax
\mciteBstWouldAddEndPuncttrue
\mciteSetBstMidEndSepPunct{\mcitedefaultmidpunct}
{\mcitedefaultendpunct}{\mcitedefaultseppunct}\relax
\EndOfBibitem
\bibitem[Mondal \latin{et~al.}(2025)Mondal, Halder, Basu, Kumar, and
  Karmakar]{mondal2025graph}
Mondal,~S.; Halder,~S.; Basu,~D.; Kumar,~S.; Karmakar,~T. Graph-Coarsening for
  Machine Learning Coarse-grained Molecular Dynamics. \emph{arXiv preprint
  arXiv:2507.16531} \textbf{2025}, \relax
\mciteBstWouldAddEndPunctfalse
\mciteSetBstMidEndSepPunct{\mcitedefaultmidpunct}
{}{\mcitedefaultseppunct}\relax
\EndOfBibitem
\bibitem[No{\'e} \latin{et~al.}(2019)No{\'e}, Olsson, K{\"o}hler, and
  Wu]{noe2019boltzmann}
No{\'e},~F.; Olsson,~S.; K{\"o}hler,~J.; Wu,~H. Boltzmann generators: Sampling
  equilibrium states of many-body systems with deep learning. \emph{Science}
  \textbf{2019}, \emph{365}, eaaw1147\relax
\mciteBstWouldAddEndPuncttrue
\mciteSetBstMidEndSepPunct{\mcitedefaultmidpunct}
{\mcitedefaultendpunct}{\mcitedefaultseppunct}\relax
\EndOfBibitem
\bibitem[Schreiner \latin{et~al.}(2023)Schreiner, Winther, and
  Olsson]{schreiner2023implicit}
Schreiner,~M.; Winther,~O.; Olsson,~S. Implicit transfer operator learning:
  Multiple time-resolution models for molecular dynamics. \emph{Advances in
  Neural Information Processing Systems} \textbf{2023}, \emph{36},
  36449--36462\relax
\mciteBstWouldAddEndPuncttrue
\mciteSetBstMidEndSepPunct{\mcitedefaultmidpunct}
{\mcitedefaultendpunct}{\mcitedefaultseppunct}\relax
\EndOfBibitem
\bibitem[Tamagnone \latin{et~al.}(2024)Tamagnone, Laio, and
  Gabri{\'e}]{tamagnoneCoarseGrainedMolecularDynamics2024a}
Tamagnone,~S.; Laio,~A.; Gabri{\'e},~M. Coarse-{{Grained Molecular Dynamics}}
  with {{Normalizing Flows}}. \emph{Journal of Chemical Theory and Computation}
  \textbf{2024}, \emph{20}, 7796--7805\relax
\mciteBstWouldAddEndPuncttrue
\mciteSetBstMidEndSepPunct{\mcitedefaultmidpunct}
{\mcitedefaultendpunct}{\mcitedefaultseppunct}\relax
\EndOfBibitem
\bibitem[Wang and Gómez-Bombarelli(2019)Wang, and
  Gómez-Bombarelli]{WangBombarelli2019}
Wang,~W.; Gómez-Bombarelli,~R. Coarse-graining auto-encoders for molecular
  dynamics. \emph{npj Computational Materials} \textbf{2019}, \emph{5}\relax
\mciteBstWouldAddEndPuncttrue
\mciteSetBstMidEndSepPunct{\mcitedefaultmidpunct}
{\mcitedefaultendpunct}{\mcitedefaultseppunct}\relax
\EndOfBibitem
\bibitem[Costa and Zavadlav(2025)Costa, and Zavadlav]{costa2025morphology}
Costa,~N.; Zavadlav,~J. Morphology-Specific Peptide Discovery via Masked
  Conditional Generative Modeling. \emph{arXiv preprint arXiv:2509.02060}
  \textbf{2025}, \relax
\mciteBstWouldAddEndPunctfalse
\mciteSetBstMidEndSepPunct{\mcitedefaultmidpunct}
{}{\mcitedefaultseppunct}\relax
\EndOfBibitem
\bibitem[Fu \latin{et~al.}(2022)Fu, Xie, Rebello, Olsen, and
  Jaakkola]{fu2022simulate}
Fu,~X.; Xie,~T.; Rebello,~N.~J.; Olsen,~B.~D.; Jaakkola,~T. Simulate
  time-integrated coarse-grained molecular dynamics with multi-scale graph
  networks. \emph{arXiv preprint arXiv:2204.10348} \textbf{2022}, \relax
\mciteBstWouldAddEndPunctfalse
\mciteSetBstMidEndSepPunct{\mcitedefaultmidpunct}
{}{\mcitedefaultseppunct}\relax
\EndOfBibitem
\bibitem[Hummerich \latin{et~al.}(2025)Hummerich, Bereau, and
  K{\"o}the]{hummerich2025split}
Hummerich,~S.; Bereau,~T.; K{\"o}the,~U. Split-Flows: Measure Transport and
  Information Loss Across Molecular Resolutions. \emph{arXiv preprint
  arXiv:2511.01464} \textbf{2025}, \relax
\mciteBstWouldAddEndPunctfalse
\mciteSetBstMidEndSepPunct{\mcitedefaultmidpunct}
{}{\mcitedefaultseppunct}\relax
\EndOfBibitem
\bibitem[Lewis \latin{et~al.}(2025)Lewis, Hempel, Jim{\'e}nez-Luna, Gastegger,
  Xie, Foong, Satorras, Abdin, Veeling, Zaporozhets, \latin{et~al.}
  others]{lewis2025scalable}
Lewis,~S.; Hempel,~T.; Jim{\'e}nez-Luna,~J.; Gastegger,~M.; Xie,~Y.;
  Foong,~A.~Y.; Satorras,~V.~G.; Abdin,~O.; Veeling,~B.~S.; Zaporozhets,~I.;
  others Scalable emulation of protein equilibrium ensembles with generative
  deep learning. \emph{Science} \textbf{2025}, eadv9817\relax
\mciteBstWouldAddEndPuncttrue
\mciteSetBstMidEndSepPunct{\mcitedefaultmidpunct}
{\mcitedefaultendpunct}{\mcitedefaultseppunct}\relax
\EndOfBibitem
\bibitem[Jing \latin{et~al.}(2024)Jing, Berger, and
  Jaakkola]{jing2024alphafold}
Jing,~B.; Berger,~B.; Jaakkola,~T. AlphaFold meets flow matching for generating
  protein ensembles. \emph{arXiv preprint arXiv:2402.04845} \textbf{2024},
  \relax
\mciteBstWouldAddEndPunctfalse
\mciteSetBstMidEndSepPunct{\mcitedefaultmidpunct}
{}{\mcitedefaultseppunct}\relax
\EndOfBibitem
\bibitem[Zheng \latin{et~al.}(2024)Zheng, He, Liu, Shi, Lu, Feng, Ju, Wang,
  Zhu, Min, \latin{et~al.} others]{zheng2024predicting}
Zheng,~S.; He,~J.; Liu,~C.; Shi,~Y.; Lu,~Z.; Feng,~W.; Ju,~F.; Wang,~J.;
  Zhu,~J.; Min,~Y.; others Predicting equilibrium distributions for molecular
  systems with deep learning. \emph{Nature Machine Intelligence} \textbf{2024},
  \emph{6}, 558--567\relax
\mciteBstWouldAddEndPuncttrue
\mciteSetBstMidEndSepPunct{\mcitedefaultmidpunct}
{\mcitedefaultendpunct}{\mcitedefaultseppunct}\relax
\EndOfBibitem
\bibitem[Kohler \latin{et~al.}(2023)Kohler, Chen, Kramer, Clementi, and
  No{\'e}]{kohler2023flow}
Kohler,~J.; Chen,~Y.; Kramer,~A.; Clementi,~C.; No{\'e},~F. Flow-matching:
  Efficient coarse-graining of molecular dynamics without forces. \emph{Journal
  of Chemical Theory and Computation} \textbf{2023}, \emph{19}, 942--952\relax
\mciteBstWouldAddEndPuncttrue
\mciteSetBstMidEndSepPunct{\mcitedefaultmidpunct}
{\mcitedefaultendpunct}{\mcitedefaultseppunct}\relax
\EndOfBibitem
\bibitem[Arts \latin{et~al.}(2023)Arts, Garcia~Satorras, Huang, Z\"{u}gner,
  Federici, Clementi, Noé, Pinsler, and van~den Berg]{Arts2023}
Arts,~M.; Garcia~Satorras,~V.; Huang,~C.-W.; Z\"{u}gner,~D.; Federici,~M.;
  Clementi,~C.; Noé,~F.; Pinsler,~R.; van~den Berg,~R. Two for One: Diffusion
  Models and Force Fields for Coarse-Grained Molecular Dynamics. \emph{Journal
  of Chemical Theory and Computation} \textbf{2023}, \emph{19},
  6151–6159\relax
\mciteBstWouldAddEndPuncttrue
\mciteSetBstMidEndSepPunct{\mcitedefaultmidpunct}
{\mcitedefaultendpunct}{\mcitedefaultseppunct}\relax
\EndOfBibitem
\bibitem[Durumeric \latin{et~al.}(2024)Durumeric, Chen, No{\'e}, and
  Clementi]{durumericLearningDataEfficient2024}
Durumeric,~A.~E.; Chen,~Y.; No{\'e},~F.; Clementi,~C. Learning data efficient
  coarse-grained molecular dynamics from forces and noise. \emph{arXiv preprint
  arXiv:2407.01286} \textbf{2024}, \relax
\mciteBstWouldAddEndPunctfalse
\mciteSetBstMidEndSepPunct{\mcitedefaultmidpunct}
{}{\mcitedefaultseppunct}\relax
\EndOfBibitem
\bibitem[Mate \latin{et~al.}(2024)Mate, Fleuret, and
  Bereau]{máté2024neuralthermodynamicintegrationfree}
Mate,~B.; Fleuret,~F.; Bereau,~T. Neural thermodynamic integration: Free
  energies from energy-based diffusion models. \emph{The Journal of Physical
  Chemistry Letters} \textbf{2024}, \emph{15}, 11395--11404\relax
\mciteBstWouldAddEndPuncttrue
\mciteSetBstMidEndSepPunct{\mcitedefaultmidpunct}
{\mcitedefaultendpunct}{\mcitedefaultseppunct}\relax
\EndOfBibitem
\bibitem[Nagel and Bereau(2025)Nagel, and
  Bereau]{nagelFokkerPlanckScoreLearning2025}
Nagel,~D.; Bereau,~T. Fokker--Planck Score Learning: Efficient Free-Energy
  Estimation under Periodic Boundary Conditions. \emph{The Journal of Physical
  Chemistry B} \textbf{2025}, \relax
\mciteBstWouldAddEndPunctfalse
\mciteSetBstMidEndSepPunct{\mcitedefaultmidpunct}
{}{\mcitedefaultseppunct}\relax
\EndOfBibitem
\bibitem[Plainer \latin{et~al.}(2025)Plainer, Wu, Klein, G{\"u}nnemann, and
  No{\'e}]{plainerConsistentSamplingSimulation2025}
Plainer,~M.; Wu,~H.; Klein,~L.; G{\"u}nnemann,~S.; No{\'e},~F. Consistent
  sampling and simulation: Molecular dynamics with energy-based diffusion
  models. \emph{arXiv preprint arXiv:2506.17139} \textbf{2025}, \relax
\mciteBstWouldAddEndPunctfalse
\mciteSetBstMidEndSepPunct{\mcitedefaultmidpunct}
{}{\mcitedefaultseppunct}\relax
\EndOfBibitem
\bibitem[Tan \latin{et~al.}(2025)Tan, Bose, Lin, Klein, Bronstein, and
  Tong]{tanScalableEquilibriumSampling2025}
Tan,~C.~B.; Bose,~A.~J.; Lin,~C.; Klein,~L.; Bronstein,~M.~M.; Tong,~A.
  Scalable equilibrium sampling with sequential boltzmann generators.
  \emph{arXiv preprint arXiv:2502.18462} \textbf{2025}, \relax
\mciteBstWouldAddEndPunctfalse
\mciteSetBstMidEndSepPunct{\mcitedefaultmidpunct}
{}{\mcitedefaultseppunct}\relax
\EndOfBibitem
\bibitem[Tan \latin{et~al.}(2025)Tan, Hassan, Klein, Syed, Beaini, Bronstein,
  Tong, and Neklyudov]{tanAmortizedSamplingTransferable2025}
Tan,~C.~B.; Hassan,~M.; Klein,~L.; Syed,~S.; Beaini,~D.; Bronstein,~M.~M.;
  Tong,~A.; Neklyudov,~K. Amortized sampling with transferable normalizing
  flows. \emph{arXiv preprint arXiv:2508.18175} \textbf{2025}, \relax
\mciteBstWouldAddEndPunctfalse
\mciteSetBstMidEndSepPunct{\mcitedefaultmidpunct}
{}{\mcitedefaultseppunct}\relax
\EndOfBibitem
\bibitem[Klein and No{\'e}(2024)Klein, and No{\'e}]{transbgs}
Klein,~L.; No{\'e},~F. Transferable boltzmann generators. \emph{Advances in
  Neural Information Processing Systems} \textbf{2024}, \emph{37},
  45281--45314\relax
\mciteBstWouldAddEndPuncttrue
\mciteSetBstMidEndSepPunct{\mcitedefaultmidpunct}
{\mcitedefaultendpunct}{\mcitedefaultseppunct}\relax
\EndOfBibitem
\bibitem[Moqvist \latin{et~al.}(2025)Moqvist, Chen, Schreiner, N{"u}ske, and
  Olsson]{moqvist2025thermodynamic}
Moqvist,~S.; Chen,~W.; Schreiner,~M.; N{"u}ske,~F.; Olsson,~S. Thermodynamic
  interpolation: A generative approach to molecular thermodynamics and
  kinetics. \emph{Journal of Chemical Theory and Computation} \textbf{2025},
  \emph{21}, 2535--2545\relax
\mciteBstWouldAddEndPuncttrue
\mciteSetBstMidEndSepPunct{\mcitedefaultmidpunct}
{\mcitedefaultendpunct}{\mcitedefaultseppunct}\relax
\EndOfBibitem
\bibitem[Schebek and Rogal(2025)Schebek, and Rogal]{schebek2025scalable}
Schebek,~M.; Rogal,~J. Scalable Boltzmann Generators for equilibrium sampling
  of large-scale materials. \emph{arXiv preprint arXiv:2509.25486}
  \textbf{2025}, \relax
\mciteBstWouldAddEndPunctfalse
\mciteSetBstMidEndSepPunct{\mcitedefaultmidpunct}
{}{\mcitedefaultseppunct}\relax
\EndOfBibitem
\bibitem[Diez \latin{et~al.}(2025)Diez, Schreiner, and
  Olsson]{diezTransferableGenerativeModels2025}
Diez,~J.~V.; Schreiner,~M.; Olsson,~S. Transferable Generative Models Bridge
  Femtosecond to Nanosecond Time-Step Molecular Dynamics. \emph{arXiv preprint
  arXiv:2510.07589} \textbf{2025}, \relax
\mciteBstWouldAddEndPunctfalse
\mciteSetBstMidEndSepPunct{\mcitedefaultmidpunct}
{}{\mcitedefaultseppunct}\relax
\EndOfBibitem
\bibitem[Nam \latin{et~al.}(2025)Nam, M{\'a}t{\'e}, Toshev, Kaniselvan,
  G{\'o}mez-Bombarelli, Chen, Wood, Liu, and Miller]{nam2025enhancing}
Nam,~J.; M{\'a}t{\'e},~B.; Toshev,~A.~P.; Kaniselvan,~M.;
  G{\'o}mez-Bombarelli,~R.; Chen,~R.~T.; Wood,~B.; Liu,~G.-H.; Miller,~B.~K.
  Enhancing Diffusion-Based Sampling with Molecular Collective Variables.
  \emph{arXiv preprint arXiv:2510.11923} \textbf{2025}, \relax
\mciteBstWouldAddEndPunctfalse
\mciteSetBstMidEndSepPunct{\mcitedefaultmidpunct}
{}{\mcitedefaultseppunct}\relax
\EndOfBibitem
\bibitem[Havens \latin{et~al.}(2025)Havens, Miller, Yan, Domingo-Enrich,
  Sriram, Wood, Levine, Hu, Amos, Karrer, \latin{et~al.}
  others]{havensAdjointSamplingHighly2025a}
Havens,~A.; Miller,~B.~K.; Yan,~B.; Domingo-Enrich,~C.; Sriram,~A.; Wood,~B.;
  Levine,~D.; Hu,~B.; Amos,~B.; Karrer,~B.; others Adjoint sampling: Highly
  scalable diffusion samplers via adjoint matching. \emph{arXiv preprint
  arXiv:2504.11713} \textbf{2025}, \relax
\mciteBstWouldAddEndPunctfalse
\mciteSetBstMidEndSepPunct{\mcitedefaultmidpunct}
{}{\mcitedefaultseppunct}\relax
\EndOfBibitem
\bibitem[Stupp and Koutsourelakis(2025)Stupp, and
  Koutsourelakis]{stuppEnergyBasedCoarseGrainingMolecular2025}
Stupp,~M.; Koutsourelakis,~P. Energy-Based Coarse-Graining in Molecular
  Dynamics: A Flow-Based Framework Without Data. \emph{arXiv preprint
  arXiv:2504.20940} \textbf{2025}, \relax
\mciteBstWouldAddEndPunctfalse
\mciteSetBstMidEndSepPunct{\mcitedefaultmidpunct}
{}{\mcitedefaultseppunct}\relax
\EndOfBibitem
\bibitem[Dern \latin{et~al.}(2025)Dern, Redl, Pfister, Kollovieh, L{\"u}dke,
  and G{\"u}nnemann]{dern2025energy}
Dern,~N.; Redl,~L.; Pfister,~S.; Kollovieh,~M.; L{\"u}dke,~D.;
  G{\"u}nnemann,~S. Energy-Weighted Flow Matching: Unlocking Continuous
  Normalizing Flows for Efficient and Scalable Boltzmann Sampling. \emph{arXiv
  preprint arXiv:2509.03726} \textbf{2025}, \relax
\mciteBstWouldAddEndPunctfalse
\mciteSetBstMidEndSepPunct{\mcitedefaultmidpunct}
{}{\mcitedefaultseppunct}\relax
\EndOfBibitem
\bibitem[Kim \latin{et~al.}(2025)Kim, Seong, Woo, Ahn, and
  Kim]{kim2025scalable}
Kim,~M.; Seong,~K.; Woo,~D.; Ahn,~S.; Kim,~M. On scalable and efficient
  training of diffusion samplers. \emph{arXiv preprint arXiv:2505.19552}
  \textbf{2025}, \relax
\mciteBstWouldAddEndPunctfalse
\mciteSetBstMidEndSepPunct{\mcitedefaultmidpunct}
{}{\mcitedefaultseppunct}\relax
\EndOfBibitem
\bibitem[Midgley \latin{et~al.}(2022)Midgley, Stimper, Simm, Sch{\"o}lkopf, and
  Hern{\'a}ndez-Lobato]{midgley2022flow}
Midgley,~L.~I.; Stimper,~V.; Simm,~G.~N.; Sch{\"o}lkopf,~B.;
  Hern{\'a}ndez-Lobato,~J.~M. Flow annealed importance sampling bootstrap.
  \emph{arXiv preprint arXiv:2208.01893} \textbf{2022}, \relax
\mciteBstWouldAddEndPunctfalse
\mciteSetBstMidEndSepPunct{\mcitedefaultmidpunct}
{}{\mcitedefaultseppunct}\relax
\EndOfBibitem
\bibitem[N\"{u}ske \latin{et~al.}(2019)N\"{u}ske, Boninsegna, and
  Clementi]{Nske2019}
N\"{u}ske,~F.; Boninsegna,~L.; Clementi,~C. Coarse-graining molecular systems
  by spectral matching. \emph{The Journal of Chemical Physics} \textbf{2019},
  \emph{151}\relax
\mciteBstWouldAddEndPuncttrue
\mciteSetBstMidEndSepPunct{\mcitedefaultmidpunct}
{\mcitedefaultendpunct}{\mcitedefaultseppunct}\relax
\EndOfBibitem
\bibitem[Yang \latin{et~al.}(2023)Yang, Templeton, Rosenberger, Bittracher,
  N{"u}ske, No{\'e}, and Clementi]{yang2023slicing}
Yang,~W.; Templeton,~C.; Rosenberger,~D.; Bittracher,~A.; N{"u}ske,~F.;
  No{\'e},~F.; Clementi,~C. Slicing and dicing: Optimal coarse-grained
  representation to preserve molecular kinetics. \emph{ACS Central Science}
  \textbf{2023}, \emph{9}, 186--196\relax
\mciteBstWouldAddEndPuncttrue
\mciteSetBstMidEndSepPunct{\mcitedefaultmidpunct}
{\mcitedefaultendpunct}{\mcitedefaultseppunct}\relax
\EndOfBibitem
\bibitem[Nateghi and N{"u}ske(2025)Nateghi, and
  N{"u}ske]{nateghi2025kinetically}
Nateghi,~V.; N{"u}ske,~F. Kinetically Consistent Coarse Graining Using
  Kernel-Based Extended Dynamic Mode Decomposition. \emph{Journal of Chemical
  Theory and Computation} \textbf{2025}, \emph{21}, 7236--7248\relax
\mciteBstWouldAddEndPuncttrue
\mciteSetBstMidEndSepPunct{\mcitedefaultmidpunct}
{\mcitedefaultendpunct}{\mcitedefaultseppunct}\relax
\EndOfBibitem
\bibitem[Carter \latin{et~al.}(1989)Carter, Ciccotti, Hynes, and
  Kapral]{Carter1989}
Carter,~E.; Ciccotti,~G.; Hynes,~J.~T.; Kapral,~R. Constrained reaction
  coordinate dynamics for the simulation of rare events. \emph{Chemical Physics
  Letters} \textbf{1989}, \emph{156}, 472–477\relax
\mciteBstWouldAddEndPuncttrue
\mciteSetBstMidEndSepPunct{\mcitedefaultmidpunct}
{\mcitedefaultendpunct}{\mcitedefaultseppunct}\relax
\EndOfBibitem
\bibitem[Ciccotti \latin{et~al.}(2005)Ciccotti, Kapral, and
  Vanden-Eijnden]{ciccotti2005blue}
Ciccotti,~G.; Kapral,~R.; Vanden-Eijnden,~E. Blue moon sampling, vectorial
  reaction coordinates, and unbiased constrained dynamics. \emph{ChemPhysChem}
  \textbf{2005}, \emph{6}, 1809--1814\relax
\mciteBstWouldAddEndPuncttrue
\mciteSetBstMidEndSepPunct{\mcitedefaultmidpunct}
{\mcitedefaultendpunct}{\mcitedefaultseppunct}\relax
\EndOfBibitem
\bibitem[Barducci \latin{et~al.}(2008)Barducci, Bussi, and
  Parrinello]{barducci2008well}
Barducci,~A.; Bussi,~G.; Parrinello,~M. Well-tempered metadynamics: a smoothly
  converging and tunable free-energy method. \emph{Physical review letters}
  \textbf{2008}, \emph{100}, 020603\relax
\mciteBstWouldAddEndPuncttrue
\mciteSetBstMidEndSepPunct{\mcitedefaultmidpunct}
{\mcitedefaultendpunct}{\mcitedefaultseppunct}\relax
\EndOfBibitem
\bibitem[Bonomi \latin{et~al.}(2009)Bonomi, Branduardi, Bussi, Camilloni,
  Provasi, Raiteri, Donadio, Marinelli, Pietrucci, Broglia, \latin{et~al.}
  others]{bonomi2009plumed}
Bonomi,~M.; Branduardi,~D.; Bussi,~G.; Camilloni,~C.; Provasi,~D.; Raiteri,~P.;
  Donadio,~D.; Marinelli,~F.; Pietrucci,~F.; Broglia,~R.~A.; others PLUMED: A
  portable plugin for free-energy calculations with molecular dynamics.
  \emph{Computer Physics Communications} \textbf{2009}, \emph{180},
  1961--1972\relax
\mciteBstWouldAddEndPuncttrue
\mciteSetBstMidEndSepPunct{\mcitedefaultmidpunct}
{\mcitedefaultendpunct}{\mcitedefaultseppunct}\relax
\EndOfBibitem
\bibitem[Laio and Gervasio(2008)Laio, and Gervasio]{laio2008metadynamics}
Laio,~A.; Gervasio,~F.~L. Metadynamics: a method to simulate rare events and
  reconstruct the free energy in biophysics, chemistry and material science.
  \emph{Reports on Progress in Physics} \textbf{2008}, \emph{71}, 126601\relax
\mciteBstWouldAddEndPuncttrue
\mciteSetBstMidEndSepPunct{\mcitedefaultmidpunct}
{\mcitedefaultendpunct}{\mcitedefaultseppunct}\relax
\EndOfBibitem
\bibitem[Tribello \latin{et~al.}(2025)Tribello, Bonomi, Bussi, Camilloni,
  Armstrong, Arsiccio, Aureli, Ballabio, Bernetti, Bonati, \latin{et~al.}
  others]{tribello2025plumed}
Tribello,~G.~A.; Bonomi,~M.; Bussi,~G.; Camilloni,~C.; Armstrong,~B.~I.;
  Arsiccio,~A.; Aureli,~S.; Ballabio,~F.; Bernetti,~M.; Bonati,~L.; others
  PLUMED Tutorials: A collaborative, community-driven learning ecosystem.
  \emph{The Journal of Chemical Physics} \textbf{2025}, \emph{162}\relax
\mciteBstWouldAddEndPuncttrue
\mciteSetBstMidEndSepPunct{\mcitedefaultmidpunct}
{\mcitedefaultendpunct}{\mcitedefaultseppunct}\relax
\EndOfBibitem
\bibitem[Batatia \latin{et~al.}(2022)Batatia, Kovacs, Simm, Ortner, and
  Cs{\'a}nyi]{batatia2022mace}
Batatia,~I.; Kovacs,~D.~P.; Simm,~G.; Ortner,~C.; Cs{\'a}nyi,~G. MACE: Higher
  order equivariant message passing neural networks for fast and accurate force
  fields. \emph{Advances in neural information processing systems}
  \textbf{2022}, \emph{35}, 11423--11436\relax
\mciteBstWouldAddEndPuncttrue
\mciteSetBstMidEndSepPunct{\mcitedefaultmidpunct}
{\mcitedefaultendpunct}{\mcitedefaultseppunct}\relax
\EndOfBibitem
\bibitem[Mirarchi \latin{et~al.}(2024)Mirarchi, Pel{\'a}ez, Simeon, and
  De~Fabritiis]{mirarchi2024amaro}
Mirarchi,~A.; Pel{\'a}ez,~R.~P.; Simeon,~G.; De~Fabritiis,~G. AMARO: All
  heavy-atom transferable neural network potentials of protein thermodynamics.
  \emph{Journal of Chemical Theory and Computation} \textbf{2024}, \emph{20},
  9871--9878\relax
\mciteBstWouldAddEndPuncttrue
\mciteSetBstMidEndSepPunct{\mcitedefaultmidpunct}
{\mcitedefaultendpunct}{\mcitedefaultseppunct}\relax
\EndOfBibitem
\bibitem[Raja \latin{et~al.}(2025)Raja, {\v{S}}{\'\i}pka, Psenka, Kreiman,
  Pavelka, and Krishnapriyan]{rajaActionMinimizationMeetsGenerative2025}
Raja,~S.; {\v{S}}{\'\i}pka,~M.; Psenka,~M.; Kreiman,~T.; Pavelka,~M.;
  Krishnapriyan,~A.~S. Action-Minimization Meets Generative Modeling: Efficient
  Transition Path Sampling with the Onsager-Machlup Functional. \emph{arXiv
  preprint arXiv:2504.18506} \textbf{2025}, \relax
\mciteBstWouldAddEndPunctfalse
\mciteSetBstMidEndSepPunct{\mcitedefaultmidpunct}
{}{\mcitedefaultseppunct}\relax
\EndOfBibitem
\bibitem[Van Der~Spoel \latin{et~al.}(2005)Van Der~Spoel, Lindahl, Hess,
  Groenhof, Mark, and Berendsen]{van2005gromacs}
Van Der~Spoel,~D.; Lindahl,~E.; Hess,~B.; Groenhof,~G.; Mark,~A.~E.;
  Berendsen,~H.~J. GROMACS: fast, flexible, and free. \emph{Journal of
  computational chemistry} \textbf{2005}, \emph{26}, 1701--1718\relax
\mciteBstWouldAddEndPuncttrue
\mciteSetBstMidEndSepPunct{\mcitedefaultmidpunct}
{\mcitedefaultendpunct}{\mcitedefaultseppunct}\relax
\EndOfBibitem
\bibitem[Fuchs \latin{et~al.}(2025)Fuchs, Chen, Thaler, and
  Zavadlav]{fuchs2025chemtraina}
Fuchs,~P.; Chen,~W.; Thaler,~S.; Zavadlav,~J. chemtrain-deploy: A parallel and
  scalable framework for machine learning potentials in million-atom MD
  simulations. \emph{Journal of Chemical Theory and Computation} \textbf{2025},
  \emph{21}, 7550--7560\relax
\mciteBstWouldAddEndPuncttrue
\mciteSetBstMidEndSepPunct{\mcitedefaultmidpunct}
{\mcitedefaultendpunct}{\mcitedefaultseppunct}\relax
\EndOfBibitem
\bibitem[Fuchs \latin{et~al.}(2025)Fuchs, Thaler, R{\"o}cken, and
  Zavadlav]{fuchs2025chemtrainb}
Fuchs,~P.; Thaler,~S.; R{\"o}cken,~S.; Zavadlav,~J. chemtrain: Learning deep
  potential models via automatic differentiation and statistical physics.
  \emph{Computer Physics Communications} \textbf{2025}, \emph{310},
  109512\relax
\mciteBstWouldAddEndPuncttrue
\mciteSetBstMidEndSepPunct{\mcitedefaultmidpunct}
{\mcitedefaultendpunct}{\mcitedefaultseppunct}\relax
\EndOfBibitem
\bibitem[Schoenholz and Cubuk(2020)Schoenholz, and Cubuk]{schoenholz2020jax}
Schoenholz,~S.; Cubuk,~E.~D. Jax md: a framework for differentiable physics.
  \emph{Advances in Neural Information Processing Systems} \textbf{2020},
  \emph{33}, 11428--11441\relax
\mciteBstWouldAddEndPuncttrue
\mciteSetBstMidEndSepPunct{\mcitedefaultmidpunct}
{\mcitedefaultendpunct}{\mcitedefaultseppunct}\relax
\EndOfBibitem
\bibitem[Wang \latin{et~al.}(2019)Wang, Ribeiro, and Tiwary]{wang2019past}
Wang,~Y.; Ribeiro,~J. M.~L.; Tiwary,~P. Past--future information bottleneck for
  sampling molecular reaction coordinate simultaneously with thermodynamics and
  kinetics. \emph{Nature communications} \textbf{2019}, \emph{10}, 3573\relax
\mciteBstWouldAddEndPuncttrue
\mciteSetBstMidEndSepPunct{\mcitedefaultmidpunct}
{\mcitedefaultendpunct}{\mcitedefaultseppunct}\relax
\EndOfBibitem
\bibitem[No{\'e} and N{"u}ske(2013)No{\'e}, and N{"u}ske]{noe2013variational}
No{\'e},~F.; N{"u}ske,~F. A variational approach to modeling slow processes in
  stochastic dynamical systems. \emph{Multiscale Modeling \& Simulation}
  \textbf{2013}, \emph{11}, 635--655\relax
\mciteBstWouldAddEndPuncttrue
\mciteSetBstMidEndSepPunct{\mcitedefaultmidpunct}
{\mcitedefaultendpunct}{\mcitedefaultseppunct}\relax
\EndOfBibitem
\bibitem[P{\'e}rez-Hern{\'a}ndez \latin{et~al.}(2013)P{\'e}rez-Hern{\'a}ndez,
  Paul, Giorgino, De~Fabritiis, and No{\'e}]{perez2013identification}
P{\'e}rez-Hern{\'a}ndez,~G.; Paul,~F.; Giorgino,~T.; De~Fabritiis,~G.;
  No{\'e},~F. Identification of slow molecular order parameters for Markov
  model construction. \emph{The Journal of chemical physics} \textbf{2013},
  \emph{139}\relax
\mciteBstWouldAddEndPuncttrue
\mciteSetBstMidEndSepPunct{\mcitedefaultmidpunct}
{\mcitedefaultendpunct}{\mcitedefaultseppunct}\relax
\EndOfBibitem
\bibitem[Zhang \latin{et~al.}(2024)Zhang, Zhang, Wu, and Wang]{zhang2024flow}
Zhang,~M.; Zhang,~Z.; Wu,~H.; Wang,~Y. Flow matching for optimal reaction
  coordinates of biomolecular systems. \emph{Journal of Chemical Theory and
  Computation} \textbf{2024}, \emph{21}, 399--412\relax
\mciteBstWouldAddEndPuncttrue
\mciteSetBstMidEndSepPunct{\mcitedefaultmidpunct}
{\mcitedefaultendpunct}{\mcitedefaultseppunct}\relax
\EndOfBibitem
\bibitem[Ribeiro \latin{et~al.}(2018)Ribeiro, Bravo, Wang, and
  Tiwary]{ribeiro2018reweighted}
Ribeiro,~J. M.~L.; Bravo,~P.; Wang,~Y.; Tiwary,~P. Reweighted autoencoded
  variational Bayes for enhanced sampling (RAVE). \emph{The Journal of chemical
  physics} \textbf{2018}, \emph{149}\relax
\mciteBstWouldAddEndPuncttrue
\mciteSetBstMidEndSepPunct{\mcitedefaultmidpunct}
{\mcitedefaultendpunct}{\mcitedefaultseppunct}\relax
\EndOfBibitem
\bibitem[Chen and Ferguson(2018)Chen, and Ferguson]{chen2018molecular}
Chen,~W.; Ferguson,~A.~L. Molecular enhanced sampling with autoencoders:
  On-the-fly collective variable discovery and accelerated free energy
  landscape exploration. \emph{Journal of computational chemistry}
  \textbf{2018}, \emph{39}, 2079--2102\relax
\mciteBstWouldAddEndPuncttrue
\mciteSetBstMidEndSepPunct{\mcitedefaultmidpunct}
{\mcitedefaultendpunct}{\mcitedefaultseppunct}\relax
\EndOfBibitem
\bibitem[Herringer \latin{et~al.}(2023)Herringer, Dasetty, Gandhi, Lee, and
  Ferguson]{herringer2023permutationally}
Herringer,~N.~S.; Dasetty,~S.; Gandhi,~D.; Lee,~J.; Ferguson,~A.~L.
  Permutationally invariant networks for enhanced sampling (PINES): Discovery
  of multimolecular and solvent-inclusive collective variables. \emph{Journal
  of Chemical Theory and Computation} \textbf{2023}, \emph{20}, 178--198\relax
\mciteBstWouldAddEndPuncttrue
\mciteSetBstMidEndSepPunct{\mcitedefaultmidpunct}
{\mcitedefaultendpunct}{\mcitedefaultseppunct}\relax
\EndOfBibitem
\bibitem[Mehdi \latin{et~al.}(2024)Mehdi, Smith, Herron, Zou, and
  Tiwary]{mehdi2024enhanced}
Mehdi,~S.; Smith,~Z.; Herron,~L.; Zou,~Z.; Tiwary,~P. Enhanced sampling with
  machine learning. \emph{Annual Review of Physical Chemistry} \textbf{2024},
  \emph{75}, 347--370\relax
\mciteBstWouldAddEndPuncttrue
\mciteSetBstMidEndSepPunct{\mcitedefaultmidpunct}
{\mcitedefaultendpunct}{\mcitedefaultseppunct}\relax
\EndOfBibitem
\bibitem[Darve and Pohorille(2001)Darve, and Pohorille]{darve2001calculating}
Darve,~E.; Pohorille,~A. Calculating free energies using average force.
  \emph{The Journal of chemical physics} \textbf{2001}, \emph{115},
  9169--9183\relax
\mciteBstWouldAddEndPuncttrue
\mciteSetBstMidEndSepPunct{\mcitedefaultmidpunct}
{\mcitedefaultendpunct}{\mcitedefaultseppunct}\relax
\EndOfBibitem
\bibitem[Gao(2008)]{gao2008self}
Gao,~Y.~Q. Self-adaptive enhanced sampling in the energy and trajectory spaces:
  Accelerated thermodynamics and kinetic calculations. \emph{The Journal of
  chemical physics} \textbf{2008}, \emph{128}\relax
\mciteBstWouldAddEndPuncttrue
\mciteSetBstMidEndSepPunct{\mcitedefaultmidpunct}
{\mcitedefaultendpunct}{\mcitedefaultseppunct}\relax
\EndOfBibitem
\bibitem[Zhang \latin{et~al.}(2018)Zhang, Wang, \latin{et~al.}
  others]{zhang2018reinforced}
Zhang,~L.; Wang,~H.; others Reinforced dynamics for enhanced sampling in large
  atomic and molecular systems. \emph{The Journal of chemical physics}
  \textbf{2018}, \emph{148}\relax
\mciteBstWouldAddEndPuncttrue
\mciteSetBstMidEndSepPunct{\mcitedefaultmidpunct}
{\mcitedefaultendpunct}{\mcitedefaultseppunct}\relax
\EndOfBibitem
\bibitem[Valsson and Parrinello(2014)Valsson, and
  Parrinello]{valsson2014variational}
Valsson,~O.; Parrinello,~M. Variational approach to enhanced sampling and free
  energy calculations. \emph{Physical review letters} \textbf{2014},
  \emph{113}, 090601\relax
\mciteBstWouldAddEndPuncttrue
\mciteSetBstMidEndSepPunct{\mcitedefaultmidpunct}
{\mcitedefaultendpunct}{\mcitedefaultseppunct}\relax
\EndOfBibitem
\bibitem[Piaggi and Parrinello(2019)Piaggi, and
  Parrinello]{piaggi2019multithermal}
Piaggi,~P.~M.; Parrinello,~M. Multithermal-multibaric molecular simulations
  from a variational principle. \emph{Physical review letters} \textbf{2019},
  \emph{122}, 050601\relax
\mciteBstWouldAddEndPuncttrue
\mciteSetBstMidEndSepPunct{\mcitedefaultmidpunct}
{\mcitedefaultendpunct}{\mcitedefaultseppunct}\relax
\EndOfBibitem
\bibitem[Wang \latin{et~al.}(2025)Wang, Wu, and Olsson]{wang2025marginal}
Wang,~Y.; Wu,~H.; Olsson,~S. Marginal Girsanov Reweighting: Stable Variance
  Reduction via Neural Ratio Estimation. \emph{arXiv preprint arXiv:2509.25872}
  \textbf{2025}, \relax
\mciteBstWouldAddEndPunctfalse
\mciteSetBstMidEndSepPunct{\mcitedefaultmidpunct}
{}{\mcitedefaultseppunct}\relax
\EndOfBibitem
\bibitem[Bause and Bereau(2021)Bause, and Bereau]{bause2021reweighting}
Bause,~M.; Bereau,~T. Reweighting non-equilibrium steady-state dynamics along
  collective variables. \emph{The Journal of Chemical Physics} \textbf{2021},
  \emph{154}\relax
\mciteBstWouldAddEndPuncttrue
\mciteSetBstMidEndSepPunct{\mcitedefaultmidpunct}
{\mcitedefaultendpunct}{\mcitedefaultseppunct}\relax
\EndOfBibitem
\bibitem[Prinz \latin{et~al.}(2011)Prinz, Wu, Sarich, Keller, Senne, Held,
  Chodera, Sch{\"u}tte, and No{\'e}]{prinz2011markov}
Prinz,~J.-H.; Wu,~H.; Sarich,~M.; Keller,~B.; Senne,~M.; Held,~M.;
  Chodera,~J.~D.; Sch{\"u}tte,~C.; No{\'e},~F. Markov models of molecular
  kinetics: Generation and validation. \emph{The Journal of chemical physics}
  \textbf{2011}, \emph{134}\relax
\mciteBstWouldAddEndPuncttrue
\mciteSetBstMidEndSepPunct{\mcitedefaultmidpunct}
{\mcitedefaultendpunct}{\mcitedefaultseppunct}\relax
\EndOfBibitem
\bibitem[Bause \latin{et~al.}(2019)Bause, Wittenstein, Kremer, and
  Bereau]{bause2019microscopic}
Bause,~M.; Wittenstein,~T.; Kremer,~K.; Bereau,~T. Microscopic reweighting for
  nonequilibrium steady-state dynamics. \emph{Physical Review E} \textbf{2019},
  \emph{100}, 060103\relax
\mciteBstWouldAddEndPuncttrue
\mciteSetBstMidEndSepPunct{\mcitedefaultmidpunct}
{\mcitedefaultendpunct}{\mcitedefaultseppunct}\relax
\EndOfBibitem
\bibitem[Krishna \latin{et~al.}(2009)Krishna, Noid, and
  Voth]{krishna2009multiscale}
Krishna,~V.; Noid,~W.~G.; Voth,~G.~A. The multiscale coarse-graining method.
  IV. Transferring coarse-grained potentials between temperatures. \emph{The
  Journal of chemical physics} \textbf{2009}, \emph{131}\relax
\mciteBstWouldAddEndPuncttrue
\mciteSetBstMidEndSepPunct{\mcitedefaultmidpunct}
{\mcitedefaultendpunct}{\mcitedefaultseppunct}\relax
\EndOfBibitem
\bibitem[Jin \latin{et~al.}(2020)Jin, Yu, and Voth]{jin2020temperature}
Jin,~J.; Yu,~A.; Voth,~G.~A. Temperature and phase transferable bottom-up
  coarse-grained models. \emph{Journal of chemical theory and computation}
  \textbf{2020}, \emph{16}, 6823--6842\relax
\mciteBstWouldAddEndPuncttrue
\mciteSetBstMidEndSepPunct{\mcitedefaultmidpunct}
{\mcitedefaultendpunct}{\mcitedefaultseppunct}\relax
\EndOfBibitem
\bibitem[Wu and Zhou(2024)Wu, and Zhou]{wu2024structural}
Wu,~Z.; Zhou,~T. Structural coarse-graining via multiobjective optimization
  with differentiable simulation. \emph{Journal of Chemical Theory and
  Computation} \textbf{2024}, \emph{20}, 2605--2617\relax
\mciteBstWouldAddEndPuncttrue
\mciteSetBstMidEndSepPunct{\mcitedefaultmidpunct}
{\mcitedefaultendpunct}{\mcitedefaultseppunct}\relax
\EndOfBibitem
\bibitem[Vitartas \latin{et~al.}(2025)Vitartas, Zhang, Jur{\'a}skov{\'a},
  Johnston-Wood, and Duarte]{vitartas2025active}
Vitartas,~V.; Zhang,~H.; Jur{\'a}skov{\'a},~V.; Johnston-Wood,~T.; Duarte,~F.
  Active learning meets metadynamics: Automated workflow for reactive machine
  learning interatomic potentials. \emph{Digital Discovery} \textbf{2025},
  \relax
\mciteBstWouldAddEndPunctfalse
\mciteSetBstMidEndSepPunct{\mcitedefaultmidpunct}
{}{\mcitedefaultseppunct}\relax
\EndOfBibitem
\bibitem[Kulichenko \latin{et~al.}(2023)Kulichenko, Barros, Lubbers, Li,
  Messerly, Tretiak, Smith, and Nebgen]{kulichenko2023uncertainty}
Kulichenko,~M.; Barros,~K.; Lubbers,~N.; Li,~Y.~W.; Messerly,~R.; Tretiak,~S.;
  Smith,~J.~S.; Nebgen,~B. Uncertainty-driven dynamics for active learning of
  interatomic potentials. \emph{Nature computational science} \textbf{2023},
  \emph{3}, 230--239\relax
\mciteBstWouldAddEndPuncttrue
\mciteSetBstMidEndSepPunct{\mcitedefaultmidpunct}
{\mcitedefaultendpunct}{\mcitedefaultseppunct}\relax
\EndOfBibitem
\bibitem[R{\"o}cken and Zavadlav(2025)R{\"o}cken, and
  Zavadlav]{rocken2025enhancing}
R{\"o}cken,~S.; Zavadlav,~J. Enhancing Machine Learning Potentials through
  Transfer Learning across Chemical Elements. \emph{Journal of Chemical
  Information and Modeling} \textbf{2025}, \emph{65}, 7406--7414\relax
\mciteBstWouldAddEndPuncttrue
\mciteSetBstMidEndSepPunct{\mcitedefaultmidpunct}
{\mcitedefaultendpunct}{\mcitedefaultseppunct}\relax
\EndOfBibitem
\bibitem[Sillitoe \latin{et~al.}(2015)Sillitoe, Lewis, Cuff, Das, Ashford,
  Dawson, Furnham, Laskowski, Lee, Lees, \latin{et~al.}
  others]{sillitoe2015cath}
Sillitoe,~I.; Lewis,~T.~E.; Cuff,~A.; Das,~S.; Ashford,~P.; Dawson,~N.~L.;
  Furnham,~N.; Laskowski,~R.~A.; Lee,~D.; Lees,~J.~G.; others CATH:
  comprehensive structural and functional annotations for genome sequences.
  \emph{Nucleic acids research} \textbf{2015}, \emph{43}, D376--D381\relax
\mciteBstWouldAddEndPuncttrue
\mciteSetBstMidEndSepPunct{\mcitedefaultmidpunct}
{\mcitedefaultendpunct}{\mcitedefaultseppunct}\relax
\EndOfBibitem
\bibitem[Mirarchi \latin{et~al.}(2024)Mirarchi, Giorgino, and
  De~Fabritiis]{mirarchi2024mdcath}
Mirarchi,~A.; Giorgino,~T.; De~Fabritiis,~G. mdCATH: A large-scale MD dataset
  for data-driven computational biophysics. \emph{Scientific Data}
  \textbf{2024}, \emph{11}, 1299\relax
\mciteBstWouldAddEndPuncttrue
\mciteSetBstMidEndSepPunct{\mcitedefaultmidpunct}
{\mcitedefaultendpunct}{\mcitedefaultseppunct}\relax
\EndOfBibitem
\bibitem[Wang \latin{et~al.}(2024)Wang, Wang, Shen, Wang, Yuan, Wu, and
  Gu]{wang2024protein}
Wang,~Y.; Wang,~L.; Shen,~Y.; Wang,~Y.; Yuan,~H.; Wu,~Y.; Gu,~Q. Protein
  conformation generation via force-guided se (3) diffusion models. \emph{arXiv
  preprint arXiv:2403.14088} \textbf{2024}, \relax
\mciteBstWouldAddEndPunctfalse
\mciteSetBstMidEndSepPunct{\mcitedefaultmidpunct}
{}{\mcitedefaultseppunct}\relax
\EndOfBibitem
\bibitem[Liu \latin{et~al.}(2025)Liu, Choi, Chen, Miller, and
  Chen]{liu2025adjoint}
Liu,~G.-H.; Choi,~J.; Chen,~Y.; Miller,~B.~K.; Chen,~R.~T. Adjoint
  Schr$\backslash$" odinger Bridge Sampler. \emph{arXiv preprint
  arXiv:2506.22565} \textbf{2025}, \relax
\mciteBstWouldAddEndPunctfalse
\mciteSetBstMidEndSepPunct{\mcitedefaultmidpunct}
{}{\mcitedefaultseppunct}\relax
\EndOfBibitem
\bibitem[He \latin{et~al.}(2025)He, Du, Vargas, Zhang, Padhy, OuYang, Gomes,
  and Hern{\'a}ndez-Lobato]{he2025no}
He,~J.; Du,~Y.; Vargas,~F.; Zhang,~D.; Padhy,~S.; OuYang,~R.; Gomes,~C.;
  Hern{\'a}ndez-Lobato,~J.~M. No Trick, No Treat: Pursuits and Challenges
  Towards Simulation-free Training of Neural Samplers. \emph{arXiv preprint
  arXiv:2502.06685} \textbf{2025}, \relax
\mciteBstWouldAddEndPunctfalse
\mciteSetBstMidEndSepPunct{\mcitedefaultmidpunct}
{}{\mcitedefaultseppunct}\relax
\EndOfBibitem
\end{mcitethebibliography}

\begin{tocentry}
    \includegraphics[width=\linewidth]{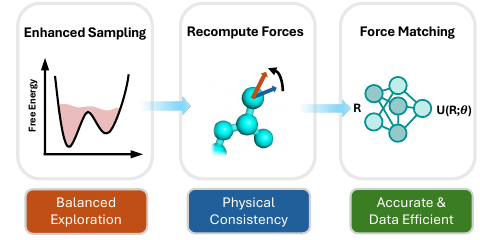}
    \label{TOC Graphic}
\end{tocentry}

\end{document}